\documentclass[preprint]{emulateapj}
\usepackage{amsmath}
\usepackage{natbib}
\usepackage{graphicx}
\usepackage{epstopdf}
\usepackage{floatrow}
\usepackage{threeparttable}
\usepackage{mathrsfs}

\newcommand{\beq}{\begin{equation}}
\newcommand{\eeq}{\end{equation}}

\begin{document}

\title{Detecting Planet Pairs in Mean Motion Resonances via Astrometry Method}
\author{Dong-Hong Wu, Hui-Gen Liu, Zhou-Yi Yu, Hui-Zhang and Ji-Lin Zhou}
\affil{School of Astronomy and Space Science, Nanjing University\\  Key Laboratory of Modern Astronomy and Astrophysics in Ministry of Education, Nanjing
University, Nanjing, 210046, China; \\
huigen@nju.edu.cn}
\begin{abstract}
GAIA leads us to step into a new era with a high astrometry precision $\sim$ 10 $\mu$as. Under such a precision, astrometry will play important roles in detecting and characterizing exoplanets. Specially, we can identify planet pairs in mean motion resonances(MMRs) via astrometry, which constrains the formation and evolution of planetary systems. In accordance with observations, we consider two Jupiters or two super-Earths systems in 1:2, 2:3 and 3:4 MMRs. Our simulations show the false alarm probabilities(FAPs) of a third planet are extremely small while the real two planets can be good fitted with signal-to-nois ratio(SNR)$>3$. The probability of reconstructing a resonant system is related with the eccentricities and resonance intensity. Generally, when SNR $\ge 10$, if eccentricities of both planets are larger than 0.01 and the resonance is quite strong, the probabilities to reconstruct the planet pair in MMRs $\ge80\%$. Jupiter pairs in MMRs are reconstructed more easily than super-Earth pairs with similar SNR when we consider the dynamical stability. FAPs are also calculated when we detect planet pairs in or near MMRs. FAPs for 1:2 MMR are largest, i.e.,  FAPs $>15\%$ when SNR $\le10$. Extrapolating from the Kepler planet pairs near MMRs and assuming SNR $\sim 3$, we will discover and reconstruct a few tens of Jupiter pairs and hundreds of super-Earth pairs in 2:3 and 1:2 MMRs within 30 pc. We also compare the differences between even and uneven data cadence and find that planets are better measured with more uniform phase coverage.
\end{abstract}

\section{Introduction}

Until April 11, 2016, 1642 planets have been detected, 1038 of them are in multiple planet systems \footnote{exoplanet.org}, and about $41\%$ of planet-host stars have more than one planetary companion. Due to the high precision of Kepler mission, many planets in multiple planet systems have been confirmed by transit timing variation(TTV) \citep{Steffen2012,Ford2012,Fabrycky2012,Xie2013,Xie2014}. However, this method is limited when two planets are very close to the resonance center, because the period of TTV is quite long and hard to be determined well. \citet{Yang2013} shows a TTV signal with period $\sim$ 1500 days based on the Kepler data as long as 1350 days. It's the longest period of TTV signal to confirm the planets near MMRs at that time. Most of the Kepler adjacent planet pairs are near or in MMRs, especially 2:3 and 1:2 MMRs \citep{Fabrycky2014,Lissauer2011,Ghilea2014,Goldreich2014}. Observations on multi-planetary systems are very important for studying the mechanisms of dynamical interaction between planets and gas disks. Many researches hint that planets end in MMRs after migration in the disk \citep{Lee2002,Papaloizou2003,Kley2004}. Additionally, planet-planet scattering may also be a major contribution to the population of resonant planets \citep{Raymond2008}. However, very few planet systems are confirmed to be in MMRs so far because of the limitation of observations. Planets detected by transit alone are lack of the information of planetary masses, while planets detected by radial velocity alone yield $m\sin{i}$. Only a few planets are detected by both transit and radial velocity methods. Besides, some of the orbital elements are degenerate and time series of planetary mean longitudes are usually not available in extrasolar systems, which make us hardly know whether they are in MMRs or not. Two of the exceptions are the  HD 82943 and HD 45364 systems. Planets detected by radial velocity around HD 82943 and HD 45364 systems are confirmed to be in 1:2 and 2:3 MMRs by dynamical stability analysis \citep{Lee2006,Correia2009}, i.e. the systems are stable only if the planets are in MMRs. However, some systems are still not confirmed to be in or near MMRs even they are very close to the resonance center, for example, the period ratio of EPIC201505350 b and c as displayed in the K2 data is 1.503514, among the closest systems to a 2:3 commensurability detected so far \citep{Armstrong2015}.

In past days, astrometric measurements with $m$as precision such as HIPPARCOS \citep{Perryman1997} does not allow the detection of exoplanets. A star with a Jupiter at 1 AU located at 30 pc has a periodic astrometric signature of about 30 $\mu$as and is hardly detected with 1 $m$as precision. However, with the improvement of the technique, many researches have shown that astrometric observations with $\mu$as-level precision are possible, such as GAIA which have been launched in 2013 \citep{Lattanzi2000,Sozzetti2001,Lattanzi2002,Lattanzi2005,Sozzetti2010} and STEP in plan \citep{chen2014}. GAIA can achieve a single-measurement astrometric error of a few tens of $\mu$as \citep{Sahlmann2015}, which is sufficient to detect a Jupiter at 1 AU around a solar-like star within 30 pc. STEP is designed to have a single-measurement astrometric precision of 1 $\mu$as, and has potential to detect habitable super-Earth around solar-like star at 30 pc. Astrometry can provide more information of the planets, including the six orbital elements and the mass of each planet, which are essential to decide whether the planet pairs are in MMRs or not.

This paper is arranged as follows. In Section 2, we describe the astrometry method in detecting exoplanets and simulation setup. In Section 3, we investigate the limits of SNR to detect the planet pairs and analysis the fitting results of the planets in our simulations. The resonance-reconstruction probabilities of planet pairs in 1:2, 2:3 and 3:4 MMRs are shown in Section 4. In Section 5, we calculate the FAPs of a detected planet system in or near MMRs. In Section 6, we estimate how many planet pairs in MMRs can be detected and reconstructed in 30 pc. The differences between even and uneven data cadences are present in Section 7. Finally, we conclude our results and discuss how to reconstruct planet pairs in MMRs better in Section 8.

\section{Detecting exoplanets by astrometry and simulation setup}

\subsection{Setups of planet pairs in MMRs}

\begin{figure}
\begin{center}
\includegraphics[width=\textwidth]{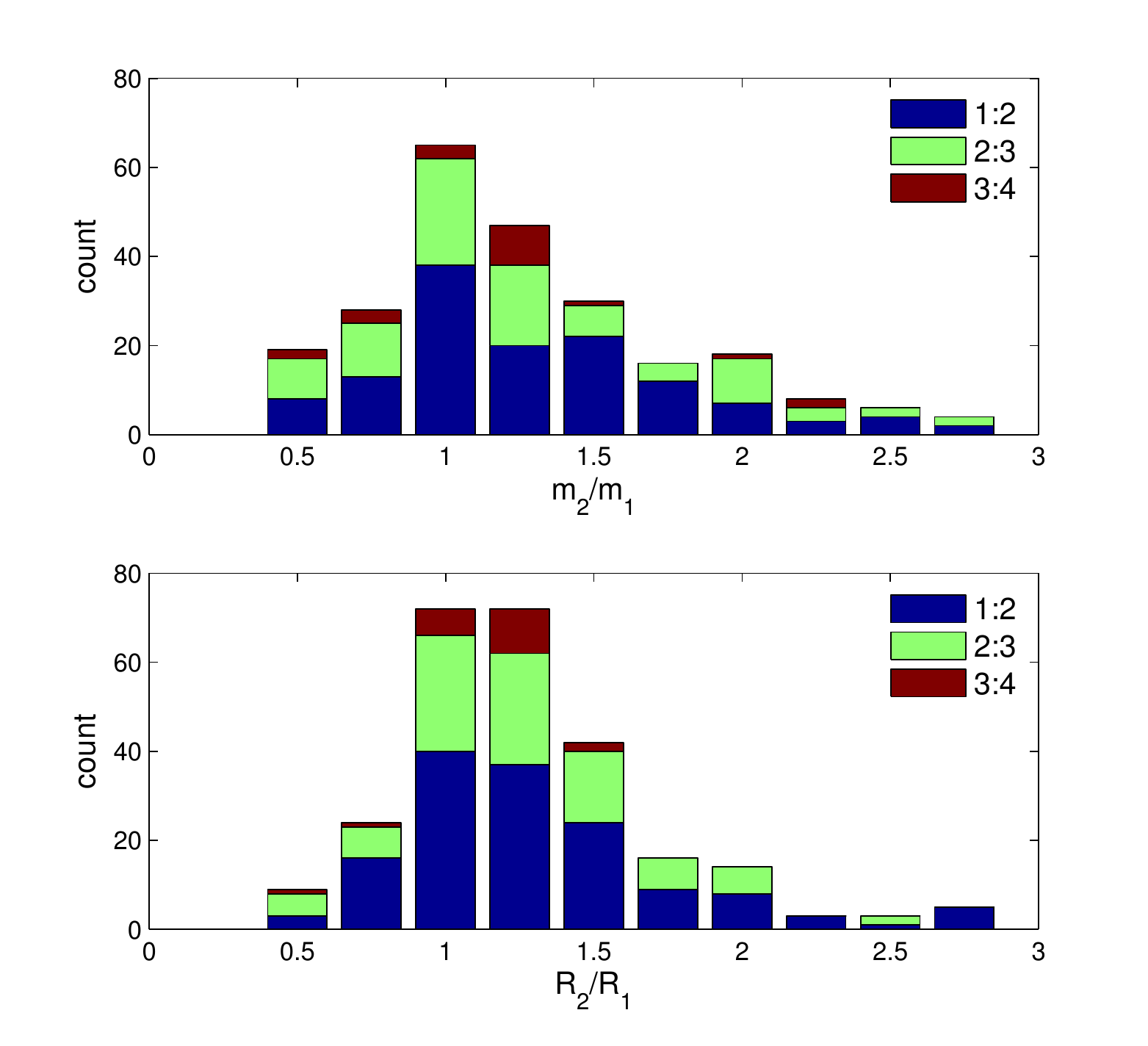}
\captionsetup{font={footnotesize}}
\ffigbox[\textwidth]{\caption{ Distribution of mass ratio $m_{2}/m_{1}$ and radius ratio $R_{2}/R_{1}$ of the near MMR planet pairs, subscript 1  and 2 represent the the inner and outer planet, respectively. The data (provided by http://exoplanet.org) includes all multiple planet systems which contain planet pairs in or near MMRs with $\Delta<0.2$($\Delta=(j-1)P_{2}/(jP_{1})-1$(j=2,3,4),$P_1$ and $P_2$ are the periods of the inner and outer planet, respectively). \label{fig1} }}.
\end{center}
\end{figure}

The mass and radius ratios of planets observed to be in or near MMRs are shown in Figure \ref{fig1}. Generally, we consider planet systems containing two planets with equal mass in 1:2, 2:3 and 3:4 MMRs in this paper. We only simulate planet systems with two Jupiters and two super-Earths separately. Both masses  of super-Earths are set as 10 Earth masses.
Hereafter, super-Earth means planet with 10 Earth masses.

Planets in MMRs can be produced by migration and randomly perturbing the orbital elements of the planets near MMRs. We simplify the migration model by adding a slow inward semi-major axis migration to the outer planet, thus the outer orbit will approach to the center of MMRs. For example, giving a number of planet pairs near $(j-1):j$ MMR ($j$=2,3,4), we add a typical migration with timescale of $5\times10^5$ years to let the planet systems evolve into MMRs. We halt the migration while the planet pairs are in $(j-1):j$ MMR ($j$=2,3,4), thus we can obtain samples of planet pairs in MMRs. With different migration time, we can attain different eccentricities of the planet pairs in MMRs. There is a positive correlation between $e_{1}$ and $e_{2}$ for resonant planet pairs produced this way. Besides, after planets migrate in the disk, they are usually locked in MMRs which are very stable \citep{Lee2009}, i.e. the resonance intensities of these systems are strong. However, if there are more planets in the disk, after the disk disappears, the resonance will be disturbed and may be not as strong as they were while migration halted. To complete our samples with different resonance intensities, we also produce resonant planet systems by randomly perturbing the orbital elements of the planets near $(j-1):j$ MMR ($j$=2,3,4). We choose planet pairs with initial $\Delta=(j-1)P_{2}/(jP_{1})-1<0.02$(j=2,3,4), $P_1$ and $P_2$ are the periods of the inner and outer planet, initial eccentricities are randomly distributed from 0 and 0.4, initial inclinations are randomly distributed from 0 to $5^\circ$, other orbital elements: $\Omega$, $\omega$ and $M$ are randomly distributed from 0 to $360^\circ$. $\Delta$ is a measure of nearness to resonance. For 2:3 and 1:2 MMRs, we have two groups of resonant planet systems produced by two methods mentioned above. But for 3:4 MMR of two Jupiters, we only have resonant systems by random method because planets are scattered before they migrate to be captured in 3:4 MMR.

\begin{figure}
\begin{center}
\includegraphics[width=\textwidth]{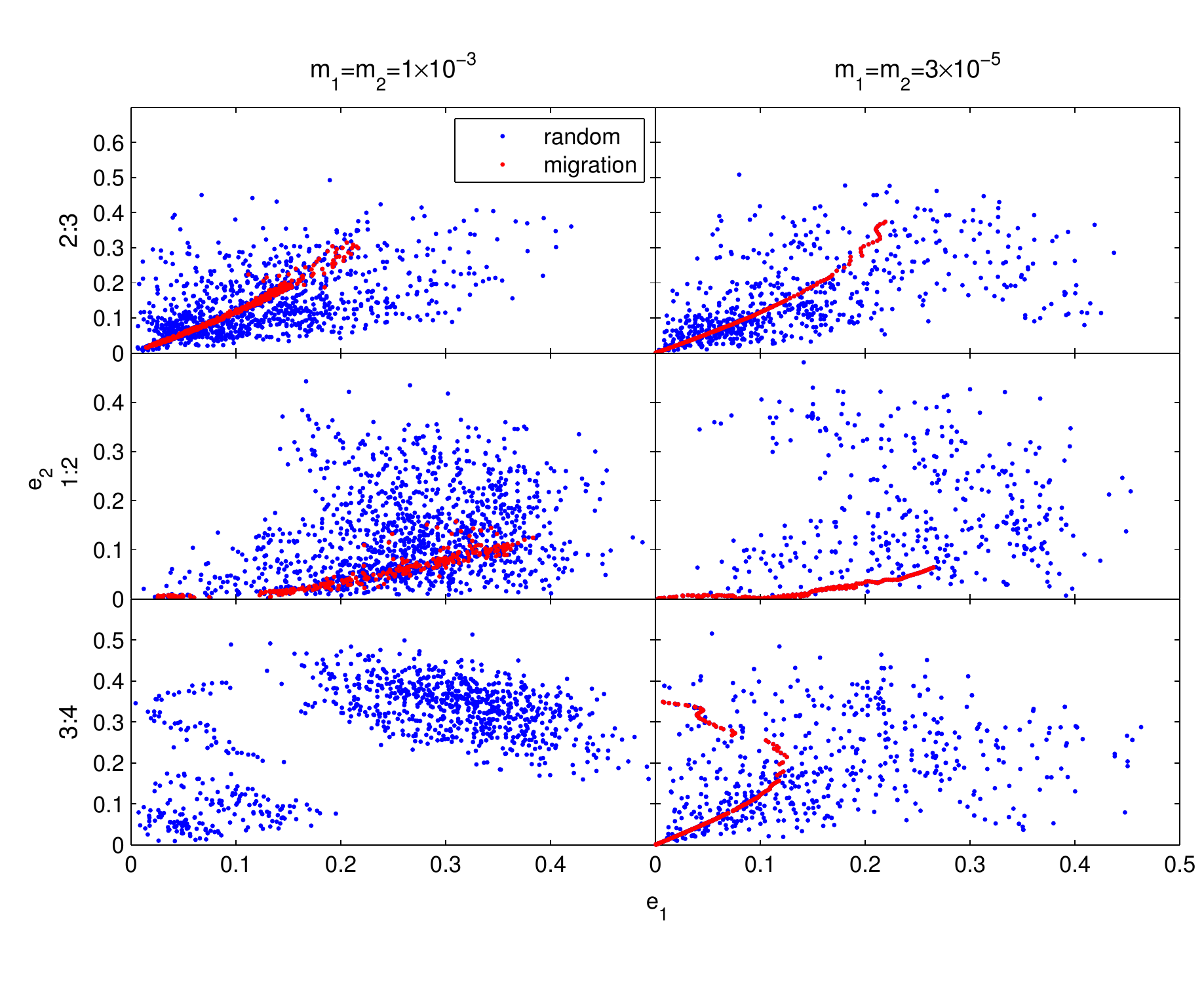}
\captionsetup{font={footnotesize}}
\caption{ Distribution of $e_{1}$ and $e_{2}$ for the samples in MMR, $e_{1}$ and $e_{2}$ are the eccentricities of the inner and outer planet, respectively. The red dots represent samples produced by migration and the blue ones represent samples produced randomly. We can see that there is a positive correlation between $e_1$ and $e_2$ for samples from migration while samples from random method have a wider distribution of eccentricities. The top, middle and the lower panels are samples of 2:3, 1:2 and 3:4 MMRs, respectively. The left panels show the samples of the Jupiter pairs and the right panels show the super-Earth pairs. \label{fig2} }.
\end{center}
\end{figure}

In this paper, all the inner planets are located at 0.8 $\sim$ 1.1 AU randomly from their host stars at 30 pc and all the planet systems are in MMRs in at least $2\times10^{4}$ years. There are only two resonance angles $\phi_{1}$ and $\phi_{2}$ for the $(j-1):j$ MMR, i.e., $\phi_{1}=j\lambda_{2}-(j-1)\lambda_{1}-\varpi_{1}$ and $\phi_{2}=j\lambda_{2}-(j-1)\lambda_{1}-\varpi_{2}$. $\lambda_{i}=M_{i}+\varpi_{i}$ is the mean longitude, $M_i$ is the mean anomaly while $\varpi_{i}=\Omega_{i}+\omega_{i}$(i=1,2).
Subscript 1 and 2 of the orbital elements represent the inner and outer planet, respectively. In general, a planet pair is considered to be in MMR as long as one resonance angle is in libration \citep{Murray1999,Raymond2008}. To obtain refined samples of planets in MMRs, we only choose planet systems with libration amplitudes of both $\phi_1$ and $\phi_2$ less than 300 degree in $2\times10^4$ years. The systems with only one resonance angle in libration are not included in our samples. The numbers of planet pairs in each MMR are shown in the first column in Table \ref{tab1} and Table \ref{tab2}. All planets in MMRs in our samples are nearly face on with inclinations between $0^\circ$ and $10^\circ$. The mutual inclinations of planet pairs are less than $5^\circ$. The eccentricity distributions are shown in Figure \ref{fig2}.

\subsection{Simulation of astrometric data}

Detecting exoplanets by astrometry with $\mu$as precision has become possible since the launch of GAIA. Similar to the RV method, astrometry measures projected movements of the host star around the barycenter of the system. By measuring the movement of the star, we can acquire planetary orbits and masses. The astrometric measurements in x and y(x and y represent the projected movement in RA and DEC direction, respectively) at time $t$ relative to the reference frame of background stars are modeled with \citep{Black1982} :
\beq
 x=x_{0}+\mu_{x}(t-t_{0})-P_{x}\pi+X+Err_{x},
 \label{x0}
 \eeq
\beq
 y=y_{0}+\mu_{y}(t-t_{0})-P_{y}\pi+Y+Err_{y},
 \label{y0}
\eeq

In Equations(\ref{x0})-(\ref{y0}), $x_{0}$ and $y_{0}$ are the coordinate offsets. $\mu_{x}$ and $\mu_{y}$ are the proper motions of the star. $P_{x}$ and $P_{y}$ are the parallax parameters which will be provided in the observation, $\pi$ is the annual parallax of the star. $x_{0}$, $y_{0}$, $\mu_{x}$, $\mu_{y}$ and $\pi$ are taken as stellar parameters. X and Y are the displacements in the star's position due to its planetary companion(s), $Err_{x}$ and $Err_{y}$ are single-measurement astrometric errors. In this paper, when we simulate astrometry data, we fix $\mu_{x}=$ 50 $m$as/year, $\mu_{y}=$ -30 $m$as/year. All the planet systems are set to be 30 pc away from us. A planet with mass $m_p$ and semi-major axis $a_p$ will lead to an astrometric signature of:
\beq
S=3(\frac{m_p}{10 m_{\rm Earth}})(\frac{a_p}{1 \rm AU})(\frac{m_{\star}}{m_{\rm Sun}})^{-1}(\frac{d}{10 \rm pc})^{-1}\mu as
\label{s}
\eeq
on the star with a distance of $d$, $m_{\star}$ is the stellar mass. We adopt a simple Gaussian measurement error model, i.e. $Err_{x}$ and $Err_{y}$ follow a Gaussian distribution with standard deviation $\sigma_m$ in our simulations.. The Signal Noise Ratio(SNR) is defined as $S/\sigma_m$, which is similar with the definition in \citet{Casertano2008}. Note, the SNR defined here is for single measurement. Equations (\ref{x0})-(\ref{y0}) can be complicated to include aberration of starlight and perspective acceleration, so we assume these effects have been perfectly removed from measurements. $P_{x}$ and $P_{y}$ can be provided given the orbit of the satellite, here we use a one-year circular orbit to simplify the parallax model. After generating planet pairs in MMRs in Section 2.1, we use a RKF7(8) \citep{Fehlberg1968} N-body code which includes the full Newtonian interaction between the planets to simulate astrometric data and sample every 0.1 year, each simulation consists of a time series of coordinate measurements according to Equations (\ref{x0})-(\ref{y0}) with a nominal mission lifetime set as 5 years. Therefore, we have a set of 50 points [$x(t_{i})$, $y(t_{i})$], $i=1,2,...,50$, each represents a measurement at observation time $t_{i}$.

\subsection{orbital parameter fitting procedure}
 In general, interaction between planets can be ignored because it can hardly affect the motion of stars \citep{Sozzetti2001}. Most of the multiple-planet systems discovered by radial velocity techniques can be well modeled by planets on independent Keplerian orbits \citep{Casertano2008}, such as the 55 Cancri system with five planets around the primary star \citep{Fischer2008}.  When a star host two planetary companions,  we also assume that the astrometric signal of the host star is the superposition of two strictly non-interacting Keplerian orbits. Ignoring the interaction between planets, X and Y are expressed as \citep{Catanzarite2010}:
\beq
X=\sum_{i=1}^N(\cos{E_{i}}-e_{i})A_{i}+\sqrt{1-e_{i}^2}(\sin{E_{i}})F_{i},
\label{x}
\eeq

\beq
Y=\sum_{i=1}^N(\cos{E_{i}}-e_{i})B_{i}+\sqrt{1-e_{i}^2}(\sin{E_{i}})G_{i}.
\label{y}
\eeq
In Equations (\ref{x})-(\ref{y}), $E_{i}$ is the eccentric anomaly, $e_{i}$ is the eccentricity of the planets, $A_{i}$, $F_{i}$, $ B_{i}$, $G_{i}$ are Thiele Innes constants, which encode amplitudes and orientations of the orbits such as the inclinations of planets $I_{i}$, arguments of pericenter $\omega_{i}$, longitudes of ascending nodes $\Omega_{i}$(i=1,...,N). N is the number of planets.

We use a hierarchical scheme to fit the orbits of the planets, the details of orbit reconstruction have been described in \citet{Catanzarite2010}, here we briefly introduce the concrete process:

 Step 1: Ignore the planetary influence on the star and invert the Equations (\ref{x0})-(\ref{y0}) by linear least squares to calculate $x_{0}$, $y_{0}$,
 $\mu_{x}$, $\mu_y$ and $\pi$, then we have the initial value of the stellar parameters for the next step.

 Step 2: Remove the coordinate offsets,
proper motion and parallax from data, and then analysis the residuals with the periodogram \citep{Scargle1982} to see if there is a significant period($P_{1}$) which exist both in x and y direction. If there is one, we obtain an initial guess of the period of the most significant planet. We identify a certain orbit when the False Alarm Probability(FAP) of the corresponding period is less than $1\%$. As we have two-dimensional time-series astrometric data, we calculate the joint periodogram defined in \citet{Catanzarite2006} as the sum of the Lomb-Scargle periodogram power from each dimension. The calculation of FAP can be found in \citet{Scargle1982} and \citet{Horne1986}.

Step 3: We randomly choose the initial value of eccentricity $e_1$ and the moment that the planet pass its perihelion $t_{01}$ of the planet. The stellar
parameters $x_0$, $y_0$, $\mu_x$, $\mu_y$, $\pi$ and the period of the planet $P_{1}$ with initial values obtained in Step 1 and 2 are also fitted.
Equations (\ref{x0})-(\ref{y0}) are easily inverted by linear least squares to yield the 4 Thiele Innes constants $A_1$, $F_1$, $B_1$ and $G_1$. X and Y can be calculated and we have fitted x and y. We adopt the MCMC algorithm in our fitting procedure. After the MCMC chains converge, we'll have more precise stellar parameters, $P_1$, $e_1$ and $t_{01}$ of the first planet.
$I_{1}$, $\omega_{1}$, $\Omega_{1}$ and $(a_1m_1)/m_{\star}$ can be calculated according to the Thiele Innes constants.

Step 4: The projected motion of star due to the first planet is then removed from the astrometric data, again we use the periodogram to search for significant peaks in the residuals. If there is one with FAP smaller than $1\%$, then it provides an initial guess for the period of the second planet($P_{2}$), the data is then fitted with a two-planet reflex motion model.

Step 5: Continue Step 2-4 until no significant signal appears in the periodogram.

For each two-planet system, there are 5+2$\times$7=19 parameters to be fitted. However, to save computing time in MCMC algorithm in Step 3 and 4, we adopt linear
equations in our fitting, only 11 parameters are fitted in the MCMC algorithm, i.e., $\mu_{x}$, $\mu_{y}$, $\pi$, $x_{0}$, $y_{0}$, $P_{1}$, $P_{2}$, $t_{01}$, $t_{02}$, $e_{1}$, $e_{2}$.
Other parameters of planets can be derived from $P_1$, $P_2$ and the 8 Thiele Innes constants. If we know the mass of the star in advance, we will also obtain semi-major axis and the mass of the each planet. Nowadays, with the development of spectrometry and astroseismology, the mass of the star can be measured with a precision of 10\% \citep{Creevey2007,Epstein2014}. Besides, semi-major axis of the planets are obtained through the relation between the mass of the host star and the orbital period in our fitting procedure, $a_{1}$ and $a_{2}$ are proportional to $m_{\star}^{1/3}$, masses of the planets $m_{1}$ and $m_{2}$ are proportional to $m_{\star}^{2/3}$, the derivation can be found in \citet{Catanzarite2010}. We briefly illustrated it here. The astrometric signature of planet $i$ has on the host star is $S_i$, with the parallax we calculated, we have an estimation of the semimajor axis of the stellar reflex motion $a_{\star,i}=S_i/\pi$(i=1,2). The center of mass equation gives the planets's mass $m_ia_i=m_{\star}a_{\star,i}$(i=1,2). Together with Kepler's $3^{rd}$ law $a_i^3=m_{\star}P_i^2$, we can determine the ratios of $a_1/a_2$ and $m_1/m_2$. Therefore, the precision of the stellar mass won't affect the characteristics of MMRs because $a_1/a_2$, $m_1/m_2$ and other orbital elements are independent of the stellar mass. As we adopt the linear function in Equations (\ref{x})-(\ref{y}), we can't distinguish the solution of parameters $\omega_{i}$, $\Omega_{i}$ from $\omega_{i}+180^\circ$, $\Omega_{i}+180^\circ$(i=1,2) without the information of position variation in the direction of our sight. However, if the orbits of planets are face on, the two solutions of parameters won't influence the resonance configuration. In this paper, to simplify the problem, we assume all the center stars have the solar mass. We run each MCMC with $3\times10^{5}$ iterations and statistics are derived on the last $1\times10^5$ elements. We choose the best-fit parameters as the median of posterior distribution. More details about the MCMC procedure can be seen in the Appendix.

\section{Detecting planets with different SNRs}

\begin{figure*}[htbp]
\centering
\includegraphics[width=\textwidth]{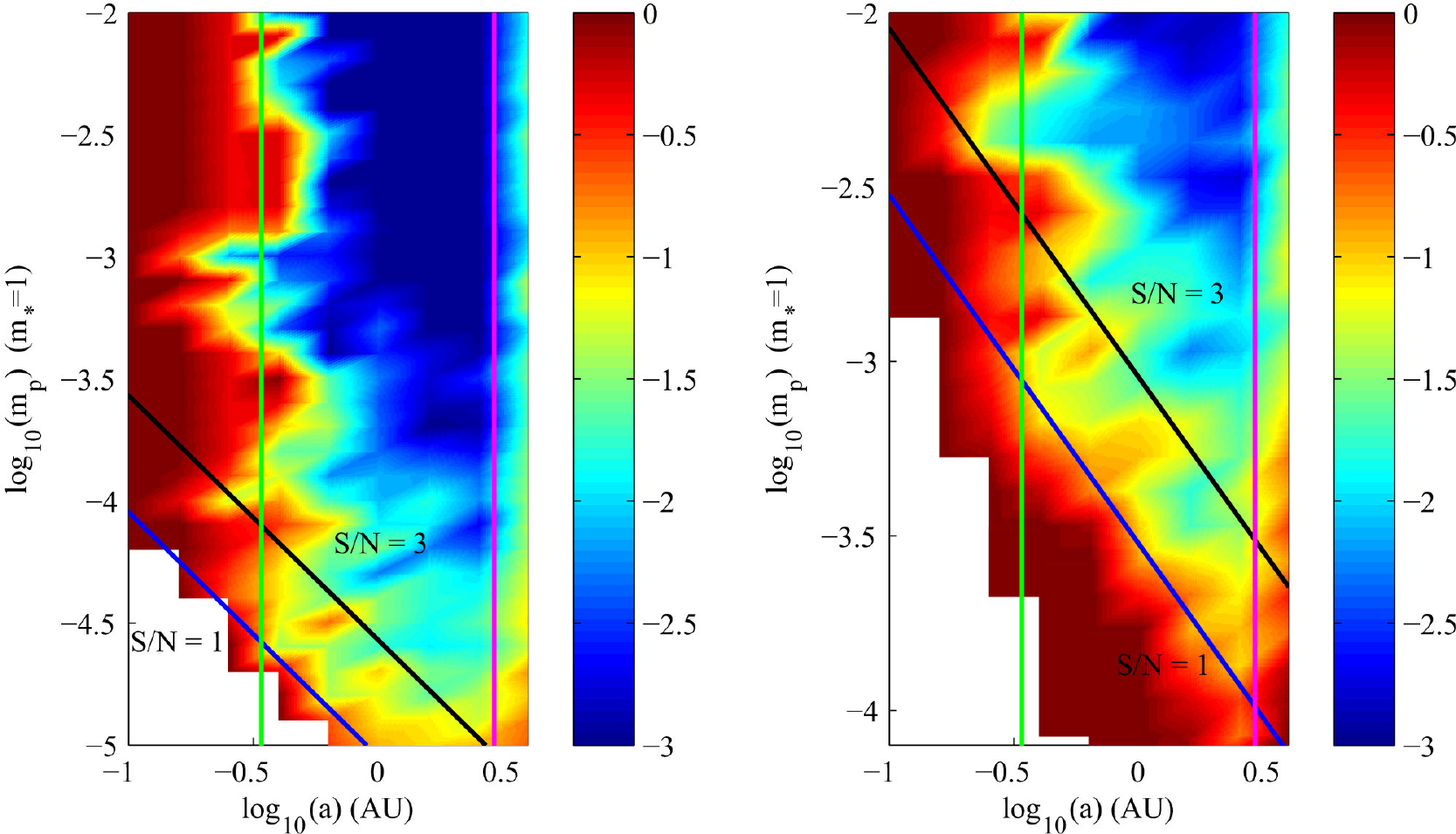}
\vspace{0cm}
\captionsetup{font={footnotesize}}
\caption{ The fitting errors $log_{10}\sqrt{(\delta_a^{2}+\delta_m^{2})/2}$ of planets with observational errors $\sigma_m=0.3$ $\mu$as(the left) and 10 $\mu$as(the right). We simulate a large number of single planet systems with different planetary mass and semi-major axis to check if we can detect and characterize them by astrometry method. All the central stars have 1 solar mass and they are 30 pc away from us. The astrometry data are generated with an even cadence of 0.1 year. The green line at $a=0.341$ AU represents a period of 0.2 year, which is the minimum period can be found with a sample cadence of 0.1 year. The magenta line at $a=2.924$ AU represents a period of 5 year. The blue line represents planet systems with SNR=1 and the dark line represents planet systems with SNR=3. The region between the green and magenta line with SNR$>$3 can be detected and characterized well. The blue regions represent planets with small relative fitting errors, the blank regions represent planets failed to be detected. \label{fig3} }.
\end{figure*}

To investegate the detection of planet with different SNR, we simulate single planet systems with different mass and semi-major axis around a solar-like star at 30 pc. We adopt a detection criterion of a planet as mentioned in section 2, i.e., the FAP of of the corresponding period is less than $1\%$. The left and right panels in Figure \ref{fig3} show the ability we can detect and characterize the planet with observational errors $\sigma_m$ $=$ 0.3 $\mu$as and 10 $\mu$as. Our simulations show planets with SNR $>$3 and period from $\sim$ 0.2 year (two times the data cadence) to 5 years (the whole observation time) can be detected reliably and consistently, which is similar with that discussed in \citet{Casertano2008}. Planets with SNR $\sim1$ can be detected but with poor determination of mass. In the worst case, planets with SNR $<1$ are hardly measured. The requirements for SNR in astrometry method is similar to that in RV method. In RV method, \citet{Cumming2004} shows that the detection of periodic signal requires $N\approx20-30$ with single signal-to-noise ratio $K/\sigma\approx2-4$ , where $K$ is the signal amplitude in radial velocity, detection of signals $<1\sigma$ requires $N\geq50$. \citet{Plavchan2015} also shows that with 50 observations, planets with signal-to-noise ratio $\ge 2$ can be detected. The lower limits of the period are due to the Nyquist sampling theorem, while the upper limits of the period are constrained by orbital phase coverage of the planets.

According to Equations \ref{s}, a star at 30 pc with a Jupiter at 1.0 AU has a periodic astrometric signature of about 30 $\mu$as, while a star at 30 pc with a super-Earth at 1.0 AU has a periodic astrometric signature of about 1 $\mu$as. As the interaction between the planets are ignored in the fitting procedure, the Jupiter pairs with observational errors $\sigma_m=10$ $\mu$as and the super-Earth pairs with observational errors $\sigma_m=0.3$ $\mu$as locate near the line SNR=3, which indicates that they can be detected and characterized well. The fitting results of two-planet system in the following sections are good, i.e. the reduced chi-square value $\chi_{\rm red}^2$ distributed between 0.9 and 1.3 for $>80\%$ cases. However, we may find a third "detection" in systems without observational error. The third detection is deduced by the Keplerian model we use. The Keplerian orbit differ from the full-Newtonian orbit, although the difference is quite small, the periodic residuals are likely to result a third detection with very small mass. Adding observational errors, there is no false detection anymore.

The fitting errors of orbital elements are shown in Table \ref{tab11} and Table \ref{tab12}. Among the six orbital elements and the planetary mass, semi-major axis are better determined than other orbital parameters. Relative fitting errors of mass are smaller than 0.06 when SNR $\ge10$. When SNR reaches 3, they can be as large as 0.13. Eccentricities can be well determined when there are not observational errors. However, when SNR $\sim3$, the absolute fitting errors of eccentricities largely increased, especially for planet pairs in 1:2 MMR. Other absolute fitting errors of orbital elements such as $I_i$, $\omega_{i}+\Omega_{i}$, $M_i$(i=1,2) are also very sensitive to observational errors. Here we compare the difference between the fitted $\omega_{i}+\Omega_{i}$ and true $\omega_{i}+\Omega_{i}$(i=1,2) because the Keplerian model we use yields two orbital solutions $\omega_i$, $\Omega_i$ and $\omega_i+180^\circ$, $\Omega_i+180^\circ$, which have been mentioned in Section 2.3. Note that as there is degeneracy between $\omega_i$ and $M_i$ when the eccentricities are very small, the absolute fitting errors of $\omega_i$ and $M_i$ decrease with the increase of eccentricities. Compared with 2:3 and 3:4 MMR, planet pairs in 1:2 MMRs have larger relative fitting errors of masses and absolute fitting errors for other orbital parameters. The 1:2 period ratio makes it hard to fit orbital parameters of both planets as well as those of the 2:3 and 3:4 MMRs because of harmonic. Average orbital parameter fitting errors of super-Earth pairs are similar with those of Jupiter pairs when they have similar SNRs. The small relative fitting errors of planetary mass and semi-major axis guarantee our successful detection and characterization of planet systems in our simulations.

\begin{table*}[htbp]
\begin{center}
\tiny
\begin{threeparttable}[b]
\captionsetup{font={footnotesize}}
\caption{average fitting errors for Jupiter pairs.}
\label{tab11}
\begin{tabular}{ccccccccc}
\tableline\tableline
&observational & $\left|a_{\rm fit}-a_{\rm true}\right|/a_{\rm true}$  &$\left|m_{\rm fit}-m_{\rm true}\right|/m_{\rm true}$ &$\left|e_{\rm fit}-e_{\rm true}\right|$&$\left|i_{\rm fit}-i_{\rm true}\right|$&$\left|\omega_{\rm fit}+\Omega_{\rm fit}-\omega_{\rm true}-\Omega_{\rm true}\right|$&$\left|M_{\rm fit}-M_{\rm true}\right|$\\\hline
2:3&0&$4.0\times10^{-4}$/$1.0\times10^{-3}$&$3\times10^{-3}$/$3.5\times10^{-3}$&$4.3\times10^{-3}$/$5.5\times10^{-3}$&
$1.6^{\circ}$/$1.8^{\circ}$&$14.7^{\circ}$/$14.0^{\circ}$&$16.9^{\circ}$/$18.6^{\circ}$\\
&3 $\mu$as&$1.2\times10^{-3}$/$1.4\times10^{-3}$&$0.024$/$0.013$&$0.03$/$0.01$&
$14.3^{\circ}$/$8.0^{\circ}$&$28.5^{\circ}$/$15.8^{\circ}$&$42.5^{\circ}$/$20.8^{\circ}$\\
&10 $\mu$as&$4.0\times10^{-3}$/$2.3\times10^{-3}$&$0.09$/$0.05$&$0.09$/$0.04$&
$28.3^{\circ}$/$16.9^{\circ}$&$60.4^{\circ}$/$39.1^{\circ}$&$85.5^{\circ}$/$49.6^{\circ}$\\\hline

1:2&0&$2.7\times10^{-4}$/$1.3\times10^{-3}$&$7.8\times10^{-3}$/$1.2\times10^{-3}$&$2.0\times10^{-3}$/$7.6\times10^{-3}$&
$0.63^{\circ}$/$0.76^{\circ}$&$4.7^{\circ}$/$19.2^{\circ}$&$35.1^{\circ}$/$25.8^{\circ}$\\
&3 $\mu$as&$9.5\times10^{-4}$/$1.5\times10^{-3}$&$0.06$/$0.01$&$0.03$/$0.07$&
$13.0^{\circ}$/$8.7^{\circ}$&$14.1^{\circ}$/$66.1^{\circ}$&$15.8^{\circ}$/$90.3^{\circ}$\\
&10 $\mu$as&$3.0\times10^{-3}$/$3.2\times10^{-3}$&$0.12$/$0.05$&$0.10$/$0.13$&
$26.2^{\circ}$/$18.2^{\circ}$&$33.8^{\circ}$/$72.3^{\circ}$&$42.3^{\circ}$/$103.2^{\circ}$\\\hline

3:4&0&$8.4\times10^{-4}$/$1.5\times10^{-3}$&$0.01$/$0.01$&$3.9\times10^{-3}$/$4.5\times10^{-3}$&
$3.1^{\circ}$/$2.8^{\circ}$&$6.1^{\circ}$/$6.9^{\circ}$&$10.0^{\circ}$/$5.0^{\circ}$\\
&3 $\mu$as&$1.3\times10^{-3}$/$1.8\times10^{-3}$&$0.014$/$0.013$&$0.019$/$0.018$&
$11.0^{\circ}$/$9.6^{\circ}$&$7.9^{\circ}$/$7.4^{\circ}$&$10.4^{\circ}$/$12.0^{\circ}$\\
&10 $\mu$as&$3.6\times10^{-3}$/$3.8\times10^{-3}$&$0.06$/$0.05$&$0.05$/$0.06$&
$21.9^{\circ}$/$19.3^{\circ}$&$18.7^{\circ}$/$16.4^{\circ}$&$26.0^{\circ}$/$27.7^{\circ}$\\\hline

\end{tabular}
  \end{threeparttable}

  \begin{flushleft}
Note. The left values near "/" are fitting errors of the inner planet while the right values are fitting errors of the outer planet.
\end{flushleft}
\end{center}
\end{table*}

\begin{table*}[htbp]
\begin{center}
\tiny
\begin{threeparttable}[b]
\captionsetup{font={footnotesize}}
\caption{average fitting errors for super-Earth pairs.}
\label{tab12}
\begin{tabular}{ccccccccc}
\tableline\tableline
&observational  & $\left|a_{\rm fit}-a_{\rm true}\right|/a_{\rm true}$ &$\left|m_{\rm fit}-m_{\rm true}\right|/m_{\rm true}$ &$\left|e_{\rm fit}-e_{\rm true}\right|$&$\left|i_{\rm fit}-i_{\rm true}\right|$&$\left|\omega_{\rm fit}+\Omega_{\rm fit}-\omega_{\rm true}-\Omega_{\rm true}\right|$&$\left|M_{\rm fit}-M_{\rm true}\right|$\\\hline
2:3&0&$3.7\times10^{-5}$/$5.7\times10^{-5}$&$3\times10^{-4}$/$2.56\times10^{-4}$&$1.3\times10^{-3}$/$9.8\times10^{-4}$&
$0.4^{\circ}$/$0.4^{\circ}$&$9.9^{\circ}$/$8.2^{\circ}$&$7.9^{\circ}$/$6.3^{\circ}$\\
&0.1 $\mu$as&$1.5\times10^{-3}$/$6.5\times10^{-4}$&$0.028$/$0.015$&$3.1\times10^{-2}$/$9.6\times10^{-3}$&
$15.1^{\circ}$/$7.4^{\circ}$&$30.9^{\circ}$/$12.8^{\circ}$&$47.9^{\circ}$/$19.3^{\circ}$\\
&0.3 $\mu$as&$4.5\times10^{-3}$/$1.8\times10^{-3}$&$0.09$/$0.05$&$0.10$/$0.03$&
$28.1^{\circ}$/$15.5^{\circ}$&$57.5^{\circ}$/$26.1^{\circ}$&$84.5^{\circ}$/$36.9^{\circ}$\\\hline

1:2&0&$9.9\times10^{-5}$/$7.0\times10^{-5}$&$0.027$/$1.2\times10^{-3}$&$8.7\times10^{-3}$/$0.016$&
$1.8^{\circ}$/$0.8^{\circ}$&$9.9^{\circ}$/$55.9^{\circ}$&$8.1^{\circ}$/$68.3^{\circ}$\\
&0.1 $\mu$as&$1.2\times10^{-3}$/$1.0\times10^{-3}$&$0.06$/$0.02$&$0.04$/$0.09$&
$15.4^{\circ}$/$7.6^{\circ}$&$17.7^{\circ}$/$68.7^{\circ}$&$27.2^{\circ}$/$87.6^{\circ}$\\
&0.3 $\mu$as&$0.017$/$3.2\times10^{-3}$&$0.13$/$0.13$&$0.13$/$0.23$&
$30.2^{\circ}$/$17.1^{\circ}$&$44.3^{\circ}$/$78.3^{\circ}$&$62.9^{\circ}$/$108.1^{\circ}$\\\hline

3:4&0&$5.7\times10^{-5}$/$7.7\times10^{-5}$&$4.2\times10^{-4}$/$3.76\times10^{-4}$&$9.6\times10^{-4}$/$9.4\times10^{-4}$&
$0.5^{\circ}$/$0.4^{\circ}$&$8.5^{\circ}$/$8.6^{\circ}$&$3.6^{\circ}$/$3.8^{\circ}$\\
&0.1 $\mu$as&$1.5\times10^{-3}$/$1.4\times10^{-3}$&$0.019$/$0.016$&$0.019$/$0.018$&
$9.7^{\circ}$/$8.6^{\circ}$&$20.9^{\circ}$/$16.6^{\circ}$&$34.9^{\circ}$/$28.8^{\circ}$\\
&0.3 $\mu$as&$4.2\times10^{-3}$/$4.2\times10^{-3}$&$0.06$/$0.05$&$0.06$/$0.06$&
$19.3^{\circ}$/$17.5^{\circ}$&$46.0^{\circ}$/$36.9^{\circ}$&$62.3^{\circ}$/$55.3^{\circ}$\\\hline

\end{tabular}
  \end{threeparttable}
  \begin{flushleft}
Note. The left values near "/" are fitting errors of the inner planet while the right values are fitting errors of the outer planet.
\end{flushleft}
\end{center}
\end{table*}

\section{The probability to reconstruct planet pairs in MMRs}

After fitting the orbital parameters of the planets, we check the stabilities of the planet systems. Because if the fitted orbital parameters deviate far from the true ones, the fitted planet systems will be unstable, especially the Jupiter pairs. We use $\beta_1$ and $\beta_2$ to indicate the probability of the fitted resonance angles $\phi_{1}$ and $\phi_{2}$ in libration. To obtain the probability of planet pairs in MMRs, we divide the total integral time $2\times10^{4}$ years into 5 equal parts and check if the resonance angles simulated in fitted systems are cycling in each $4\times10^{3}$ years. The probability of planet pairs in MMRs is defined as the fraction of time with librating resonance angle. We use $\beta$, the larger one between $\beta_1$ and $\beta_2$ to represent the probability of reconstructing a planet pair in MMR.

\subsection{The stabilities and probabilities in MMRs of the fitted planet systems}

For a two-planet system, the separation of the planets should be at least 3.5 $R_{H}$ according to Gladman1993 if the planets are Hill stable. $R_{H}$ is the Hill radius of a planet. For a Jupiter at 1 AU, 3.5 $R_{H}$ is about 0.242 AU, so the outer planet should be outside of 1.242 AU with a period ratio $P_1/P_2$ smaller than 0.72. The Hill stability indicates Jupiter pairs near 1:2 and 2:3 MMRs are likely to be stable, while those near 3:4 MMR are always unstable unless they are exactly in 3:4 MMR. This analysis is corresponding with our simulations that Jupiter pairs with $P_{1}/P_{2}\sim$ $3/4$ are stable only if they are in 3:4 MMR. The 3.5 $R_{H}$ for a super-Earth at 1 AU is 0.075 AU, so the outer planet should be outside of 1.075 AU with period ratio $P_1/P_2<1.114$, which indicates that super-Earth pairs near 2:3, 1:2 and 3:4 MMRs are most likely to be Hill stable. From observations of RV and Kepler data, the occurrence rate of Super-Earths is higher than that of Jupiters \citep{Winn2015}. Besides, for planet pairs near MMRs\footnote{exoplanet data used here are from exoplanets.org}, the fraction of both planets with $5 M_{\rm Earth} \sim 20 M_{\rm Earth}$ are $23.69\%$, the fraction of both planets with masses $0.5 M_{\rm J} \sim 2 M_{\rm J}$ are $2.79\%$. So super-Earths near MMRs are more common than two Jupiters near MMRs, especially in Kepler planet systems.

We develop an N-body code based on the RKF7(8)\citep{Fehlberg1968} integrator which includes full Newtonian interactions between the planets to check if the fitted planet systems are stable in $2\times10^{4}$ years. The stable fractions of the fitted planet systems and the fractions of stable planet systems with $\beta>0.5$ for the Jupiter pairs and super-Earth pairs are shown in Table \ref{tab1} and Table \ref{tab2}. As we use the Keplerian motion to model the true motion, even when not considering observational errors, the planet systems can not be perfectly fitted and reconstructed. Besides, when eccentricities of the planets are very small, the orbits of the planets are circular, there is geometrical degeneracy of $\omega_{i}$ and $M_{i}$(i=1,2) which makes it hard to determine $\omega_{i}$ and $M_{i}$(i=1,2) correctly. In our simulations, super-Earth pairs in 2:3 and 3:4 MMRs with $\beta<0.5$ mostly have eccentricities smaller than $5\times10^{-3}$ when not considering observational errors. For 1:2 MMRs, the 1:2 period ratio makes it hard to have good fitting results because of the influences of harmonic. Due to the reasons above, a small fraction of planet systems are not well-reconstructed.

In Table \ref{tab1} we can see that when $\sigma_m\le10$ $\mu$as, more than $90\%$ of the fitted Jupiter pairs in 2:3 and 1:2 MMRs are stable. For Jupiter pairs in 3:4 MMR, even without observational errors, only half of fitted planet systems are stable. Although the fitting errors of the Jupiter pairs in 3:4 MMR are similar to those in 2:3 MMR, it's harder for planets to be locked in 3:4 MMRs than in 2:3 MMRs, therefore, stable fractions of Jupiter pairs in 3:4 MMR are much less than those in 2:3 MMR. When only considering the MMR-reconstruction probabilities in stable fitted systems, more than $80\%$ of Jupiter pairs in 3:4 MMR can be reconstructed with $\beta>0.5$. We check the long time stabilities of a few systems in 3:4 MMR with $\beta<0.5$ and find all these systems are unstable in 0.5 Myr. Consequently, with a longer stability checking time, Jupiter pairs with low probabilities in 3:4 MMR can be excluded in the stable samples, thus the MMR-reconstruction probabilities in stable fitted systems could also approach to $100\%$.

\begin{table}[htbp]
\begin{center}
\footnotesize
\begin{threeparttable}[b]
\captionsetup{font={footnotesize}}
\caption{Fraction of stable and well-reconstructed Jupiter pairs with different observational errors.}
\label{tab1}
\begin{tabular}{cccc}
\tableline\tableline
&&\multicolumn{2}{c}{even cadence}\\\hline
&observational & fraction of stable &fraction of stable planet \\
&error&systems&systems with $\beta>0.5$\tnote{1}\\\hline
2:3&0&$95\%$&$95\%\pm1\%$\\
 (1319)\tnote{2}&3 $\mu$as&$95\%$&$87\%\pm1\%$\\
       &10 $\mu$as&$90\%$&$58\%\pm1\%$\\\hline
1:2&0&$99\%$&$98\%\pm1\%$\\
(1370)\tnote{3}&3 $\mu$as&$99\%$&$71\%\pm3\%$\\
      &10 $\mu$as&$91\%$&$58\%\pm3\%$\\\hline
3:4&0&$49\%$&$99\%\pm1\%$\\
(926)\tnote{4}&3 $\mu$as&$48\%$&$98\%\pm1\%$\\
&10 $\mu$as&$42\%$&$85\%\pm1\%$\\\hline
\end{tabular}
\begin{tablenotes}
    \item [1] the MMR-reconstruction prabability
    \item [2] Sample number of Jupiter pairs in 2:3 MMR
	\item [3] Sample number of Jupiter pairs in 1:2 MMR
    \item [4] Sample number of Jupiter pairs in 1:2 MMR
   \end{tablenotes}
  \end{threeparttable}

\begin{flushleft}
Note. The uncertainties are calculated as the difference between fraction of $\beta>0.5$ from all stable systems and fraction of $\beta>0.5$ from $N/2$ stable systems. $N$ is the sample number of each MMR shown in the parenthesis. $N/2$ samples are chosen randomly to guarantee the $N/2$ samples have similar distribution of eccentricities and $\Delta$ of the whole samples.
\end{flushleft}
\end{center}
\end{table}

For super-Earth pairs, results in Table \ref{tab2} shows that the stable fractions of fitted planet systems are generally much larger than those of the Jupiter pairs. As we have mentioned above, if two Jupiters are not in MMR, they are likely to be unstable according to Hill stability. In observations, planets with Jupiter mass observed to be near MMR are usually confirmed to be in MMR according to their dynamical stability \citep{Lee2006,Correia2009}. In this paper, we didn't do such kind of research. So when considering the MMR-reconstruction probabilities in stable fitted systems, the fractions of super-Earth pairs with $\beta>0.5$ are smaller than those of the Jupiter pairs. Considering the similar fitting errors of orbital parameters with similar SNR (Table \ref{tab11} and \ref{tab12}) and the fact that resonance width increase with planetary mass \citep{Deck2013}, stable Jupiter pairs are more likely to stay in MMRs in our reconstruction. The fraction of fitted planet systems with $\beta>0.5$ are larger than $70\%$ when SNR$\ge$10. When SNR=3, the fractions largely drop to $40-60\%$. We'll investigate the relation between the MMR-reconstruction probability in stable fitted systems and $\Delta$, eccentricity and resonance intensity in the following sections.

\begin{table}[htbp]
\begin{center}
\footnotesize
\begin{threeparttable}[b]
\captionsetup{font={footnotesize}}
\caption{ Fraction of stable and well-reconstructed super-Earth pairs with different observational errors.}
\label{tab2}
\begin{tabular}{cccc}
\tableline\tableline
&&\multicolumn{2}{c}{even cadence}\\\hline
&observational & fraction of stable &fraction of stable planet\\
&error&systems&systems with $\beta>0.5$\\\hline
2:3&0&$99\%$&$93\%\pm1\%$\\
 (812)\tnote{1}&0.1 $\mu$as&$99\%$&$76\%\pm2\%$\\
       &0.3 $\mu$as&$99\%$&$42\%\pm1\%$\\\hline
1:2&0&$100\%$&$91\%\pm1\%$\\
(562)\tnote{2}&0.1 $\mu$as&$100\%$&$79\%\pm1\%$\\
      &0.3 $\mu$as&$100\%$&$49\%\pm2\%$\\\hline
3:4&0&$99\%$&$92\%\pm1\%$\\
(895)\tnote{3}&0.1 $\mu$as&$99\%$&$77\%\pm2\%$\\
&0.3 $\mu$as&$98\%$&$40\%\pm1\%$\\\hline

\end{tabular}
\begin{tablenotes}
    \item [1] Sample number of super-Earth pairs in 2:3 MMR
	\item [2] Sample number of super-Earth pairs in 1:2 MMR
    \item [3] Sample number of super-Earth pairs in 1:2 MMR
   \end{tablenotes}
  \end{threeparttable}
  \begin{flushleft}
Note. The uncertainties are calculated similar to those in Table \ref{tab1}.
\end{flushleft}
\end{center}
\end{table}

\subsection{MMR-reconstruction with different $\Delta$}

\begin{figure}
\centering
\includegraphics[width=\textwidth]{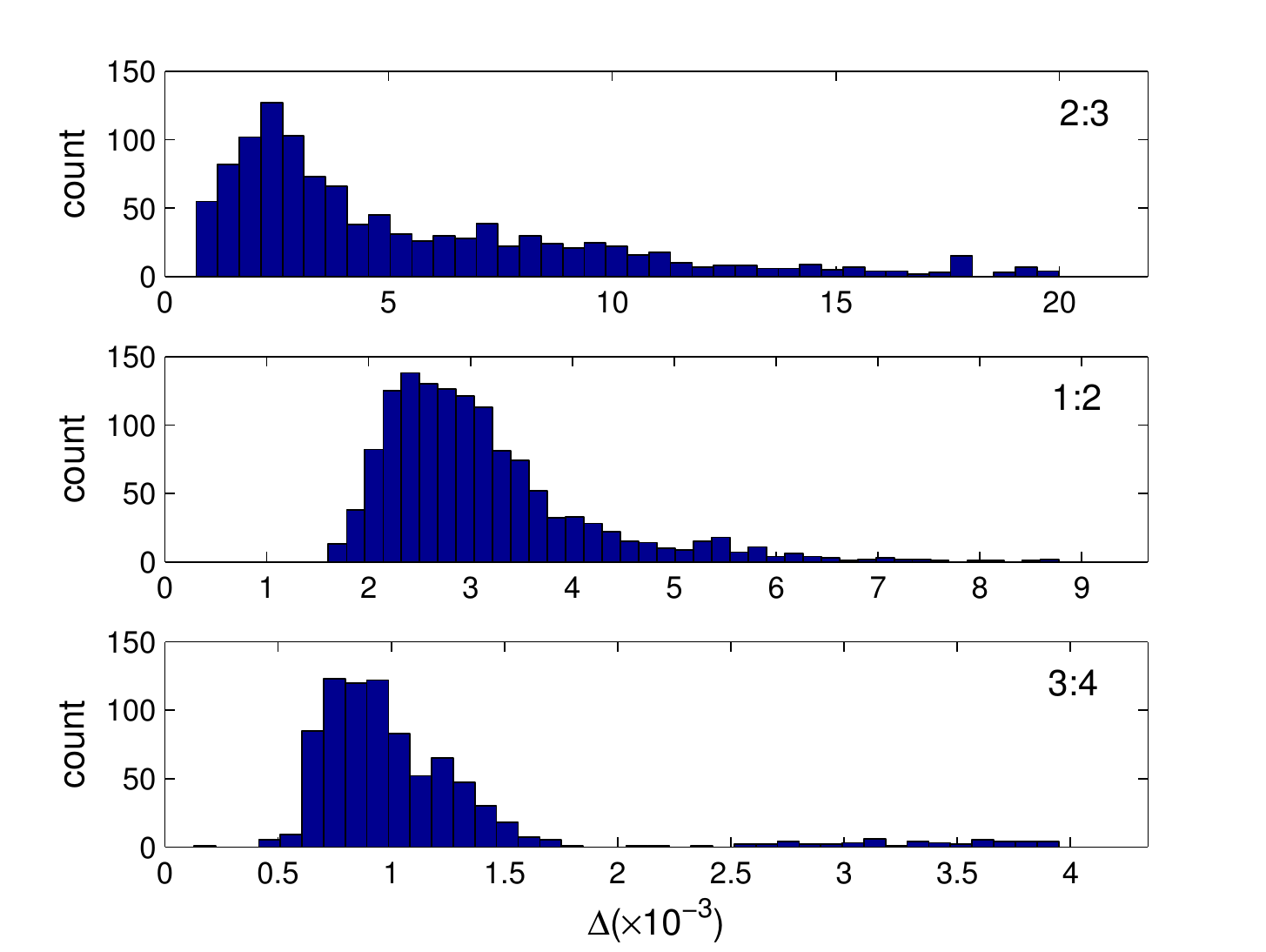}
\vspace{0cm}
\captionsetup{font={footnotesize}}
\caption{ Distribution of the normalized distance $\Delta$ from the resonant center of the Jupiter pairs. The top, middle and the lower panels are samples of the 2:3, 1:2 and 3:4 MMRs, respectively. Samples with large $\Delta$ are not shown here.  \label{fig4} }.
\end{figure}

\begin{figure}
\centering
\includegraphics[width=\textwidth]{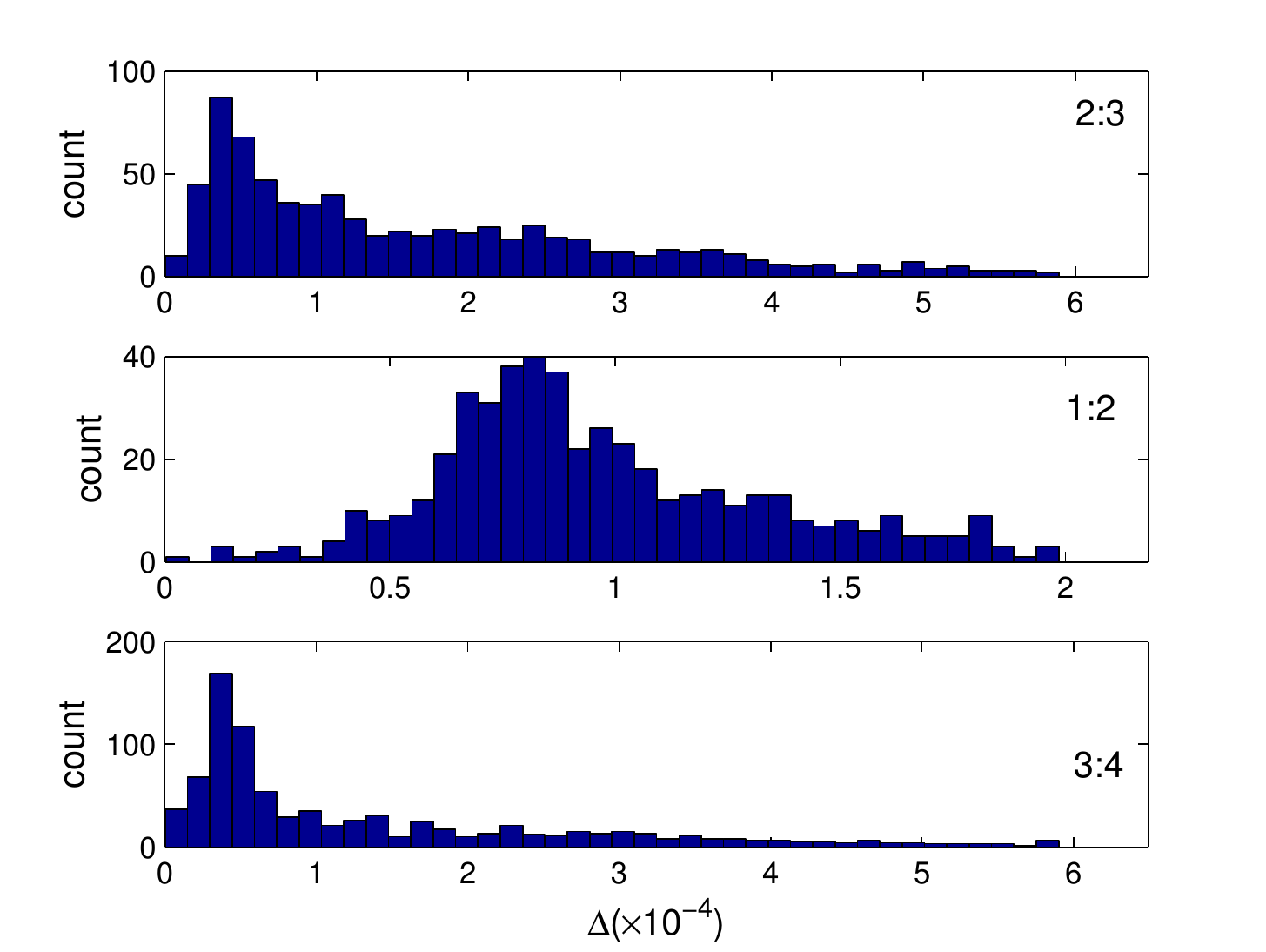}
\vspace{0cm}
\captionsetup{font={footnotesize}}
\caption{ distribution of the normalized distance $\Delta$ from the resonant center  of the super-Earth pairs. The top, middle and the lower panels are samples of the 2:3, 1:2 and 3:4 MMRs. Samples with large $\Delta$ are not shown here. \label{fig5} }.
\end{figure}

The distributions of $\Delta$ for the Jupiter pairs and super-Earth pairs in our samples are shown in Figure \ref{fig4} and Figure \ref{fig5}. We can see that $\Delta$ concentrate upon small values. For Jupiter pairs, $\Delta\sim10^{-3}$, while for super-Earth pairs, $\Delta\sim10^{-4}$. As the resonance width increased with the mass of the planet pairs \citep{Deck2013}, the values of $\Delta$ for Jupiters are much larger than super-Earths. Planet pairs in 2:3 MMR have a much wider distribution of $\Delta$ than those in 1:2 and 3:4 MMRs. Because in our simulations, planet pairs with large $\Delta$ are generated by migration, so Jupiter pairs in 3:4 MMR tend to have a small $\Delta$ for a lack of samples from migration. As we used a simplified migration model, planet pairs with small eccentricities usually have large $\Delta$. From Figure \ref{fig2} we can find that planet pairs from migration in 2:3 MMR have more samples with small eccentricities than planet pairs in 1:2 and 3:4 MMRs, so $\Delta$ distribution is broader for 2:3 MMR pairs than for 1:2 or 3:4 MMR. In this section, we will check the relation between $\Delta$ and the MMR-reconstruction probabilities.

Although relative fitting errors of semi-major axis of the planets are very small(Table \ref{tab11} and \ref{tab12}), the absolute fitting errors of $\Delta$ can be large. As $\Delta$ is calculated according to average periods of the planet pairs in $2\times10^4$ years, a small variation on initial semi-major axis will lead to large difference in $\Delta$. We calculate $\Delta_{\rm fit}$ according to average fitted periods of the planets in $2\times10^4$ years and find that the average differences between $\Delta$ and $\Delta_{\rm fit}$ are around $2\times10^{-4}$ without observational errors for both Jupiter and super-Earth pairs. When there are observational errors, the average differences between $\Delta$ and $\Delta_{\rm fit}$ reach $10^{-3}$.

\begin{figure}
\centering
\includegraphics[width=\textwidth]{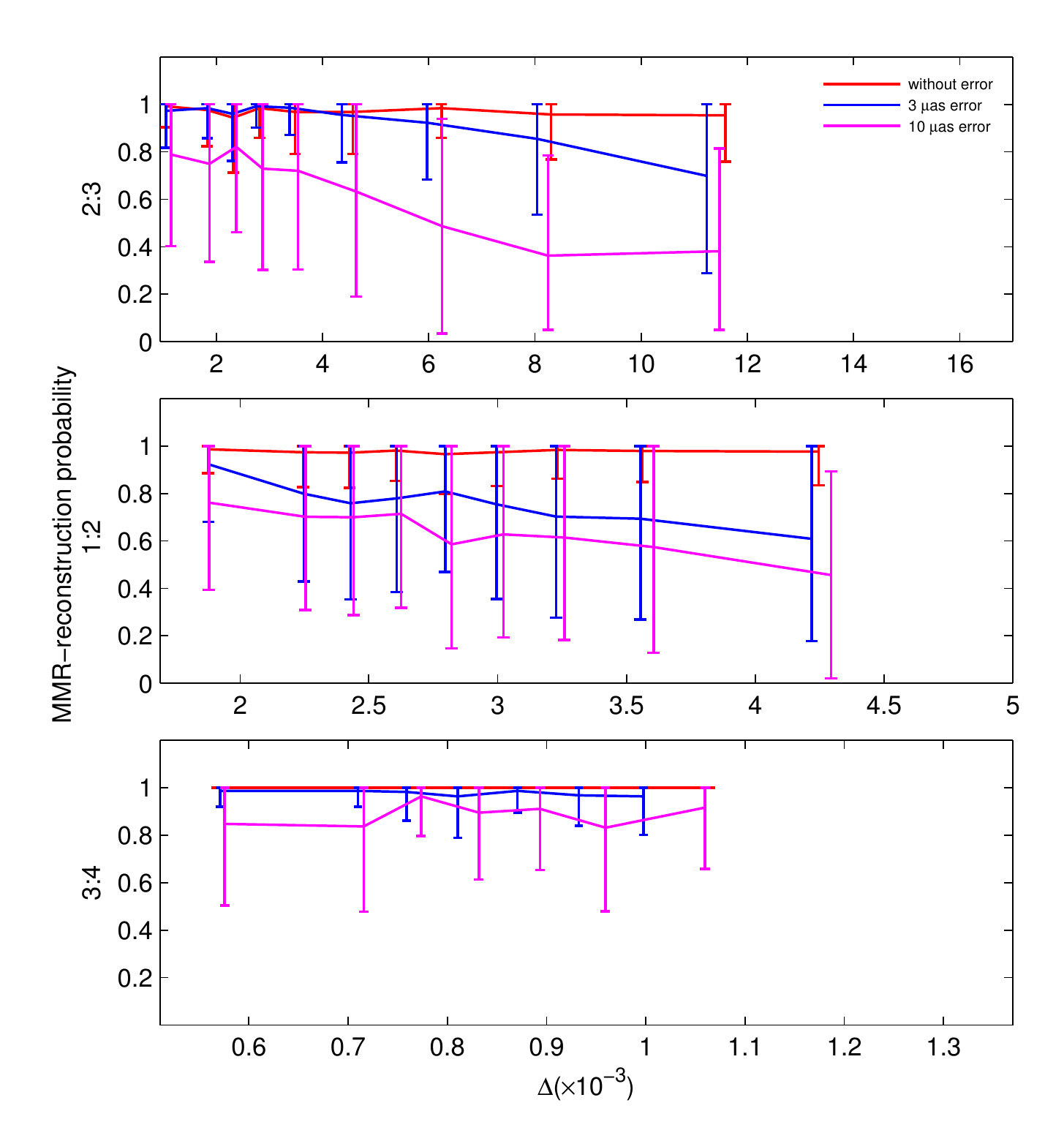}
\vspace{0cm}
\captionsetup{font={footnotesize}}
\caption{ Relations between the MMR-reconstruction probabilities of the Jupiter pairs in MMR and $\Delta$. The top, middle and the lower panels are results of the 2:3, 1:2 and 3:4 MMRs. The red, blue and magenta lines show results with observational errors with $\sigma_m=$0, 3 $\mu$as, 10 $\mu$as, respectively. \label{fig6} }.
\end{figure}

\begin{figure}
\centering
\includegraphics[width=\textwidth]{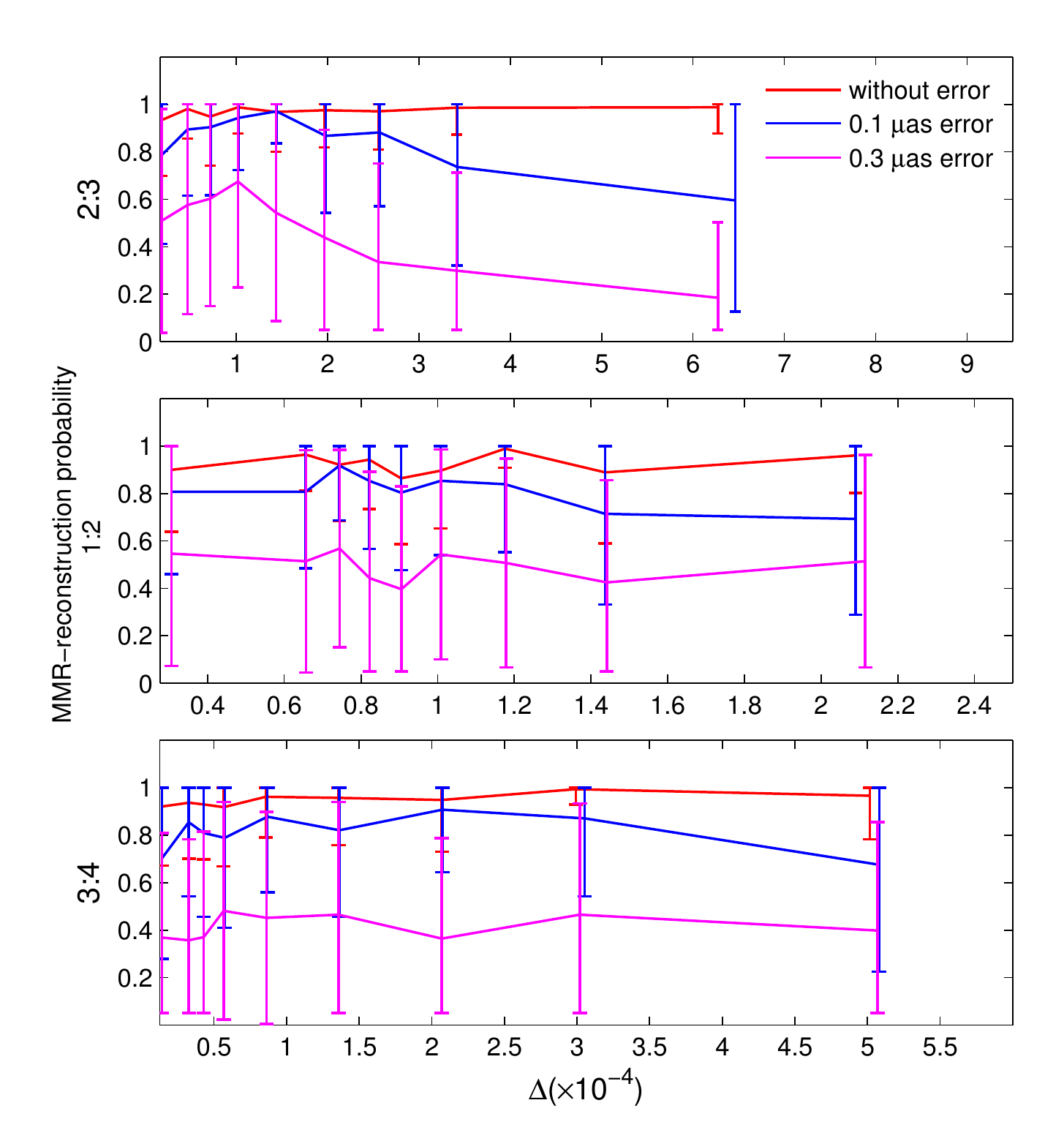}
\vspace{0cm}
\captionsetup{font={footnotesize}}
\caption{Relations between the MMR-reconstruction probabilities of the super-Earth pairs in MMR and $\Delta$. The top, middle and the lower panels are results of the 2:3, 1:2 and 3:4 MMRs. The red, blue and magenta lines show results with observational errors 0, 0.1 $\mu$as, 0.3 $\mu$as, respectively. \label{fig7} }.
\end{figure}

To check the correlations between $\Delta$ and MMR-reconstruction probability, we sort the samples in each MMR with increasing $\Delta$ and divide them into 10 parts with  the same number of samples. Define $\overline{\beta}(\Delta)$ as the average value of $\beta$ for planets in each part. Figure \ref{fig6} and Figure \ref{fig7} shows $\overline{\beta}(\Delta)$ at different $\Delta$. We don't show the $10_{th}$ part with the largest $\Delta$ for extreme large variations. If observations are carried out without any errors, we can reconstruct nearly all the systems in MMRs independent of $\Delta$.

For Jupiter pairs in 2:3 and 1:2 MMR, there is a decrease of $\overline{\beta}(\Delta)$ with the increase of $\Delta$ . With the increase of $\Delta$, the resonance becomes fragile and small variation of $\Delta$ may destroy the resonance, therefore the MMR-reconstruction probabilities decrease. If the $\sigma_m=$ 3 $\mu$as, the MMR-reconstruction probabilities are larger than $60\%$, with a slight decrease with $\Delta$. While observational errors are large enough, e.g. $\sigma_m=10$ $\mu$as, $\overline{\beta}(\Delta)$ are less than 80$\%$ for both the 2:3 and 1:2 MMRs with large dispersions. $\overline{\beta}(\Delta)$ of the 3:4 MMR are mostly constrained by stability and still $>80\%$.

For super-Earth pairs, similar with Jupiter pairs,  $\overline{\beta}(\Delta)$ should decrease with the increase of $\Delta$. However there is no obvious negative correlation. This is because of the large absolute fitting errors of $\Delta$ for super-Earth pairs. As we have mentioned in the previous paragraph, the average differences between $\Delta$ and $\Delta_{\rm fit}$ reach $10^{-3}$, which is much larger than the distribution range of $\Delta$ for Super-Earth pairs, but smaller than that for Jupiter pairs. So there is no positive correlation between $\Delta$ and $\Delta_{\rm fit}$ for super-Earth, while $\Delta_{\rm fit}$ increases with $\Delta$ for Jupiter pairs. i.e., the negative correlation between $\Delta$ and $\overline{\beta}(\Delta)$ are hidden by large fitting errors of $\Delta$ for Super Earth. $\overline{\beta}(\Delta)$ are smaller than those of the Jupiter pairs with similar SNRs. As we calculate $\overline{\beta}(\Delta)$ in the stable systems, Jupiter pairs remain stable are more likely to be in MMR because of Hill stability. When $\sigma_m=0.1$ $\mu$as, $\overline{\beta}(\Delta)$ for 2:3 MMR drops from $80\%$ to $60\%$ with the increase of $\Delta$. For 1:2 and 3:4 MMRs, $\overline{\beta}(\Delta)$ is about $80\%$. When $\sigma_m=0.3$ $\mu$as, $\overline{\beta}(\Delta)$ is smaller than $60\%$.

The large dispersions makes the relation between $\overline{\beta}(\Delta)$  and $\Delta$ a little obscure, which also indicates some other factors can influence the MMR-reconstruction probabilities such as the eccentricity and resonance intensity. We'll analysis the relations between MMR-reconstruction and eccentricity and resonance intensity in the following subsections.

\subsection{MMR-reconstruction with different eccentricities}

\begin{figure}
\centering
\includegraphics[width=\textwidth]{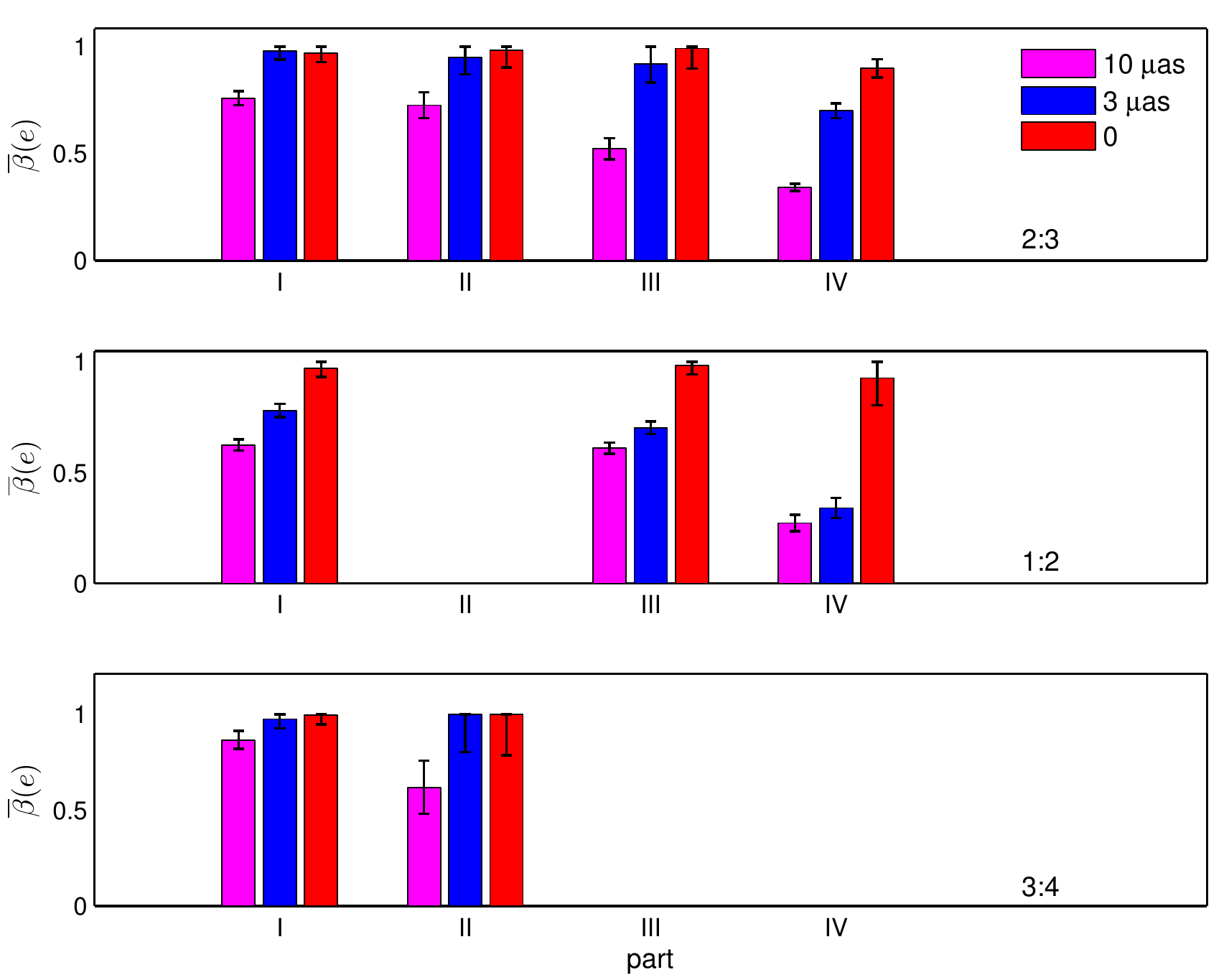}
\vspace{0cm}
\captionsetup{font={footnotesize}}
\caption{ Relations between the average resonance-reconstruction probabilities $\overline{\beta}(e)$ and eccentricities of the Jupiter pairs in MMRs. The top, middle and the lower bar graphs are results of the 2:3, 1:2 and 3:4 MMRs, respectively. The red, blue and magenta colors represent observational errors $=$0, 3 $\mu$as, 10 $\mu$as, respectively. I,II, III and IV represent different ranges of eccentricities, i.e., I: $e_{1}>0.1$, $e_{2}>0.1$, II: $e_{1}<0.1$, $e_{2}>0.1$, III: $e_{1}>0.1$, $e_{2}<0.1$, IV: $e_{1}<0.1$, $e_{2}<0.1$. The number of samples in each part is larger than 20. The error bars displayed on the bins represent only the uncertainty due to Poisson statistics. For 1:2 MMR, there are few samples in part II. For 3:4 MMR, there are few samples in part III and IV, therefore, we didn't show the results in these part here. \label{fig8} }.
\end{figure}

\begin{figure}
\centering
\includegraphics[width=\textwidth]{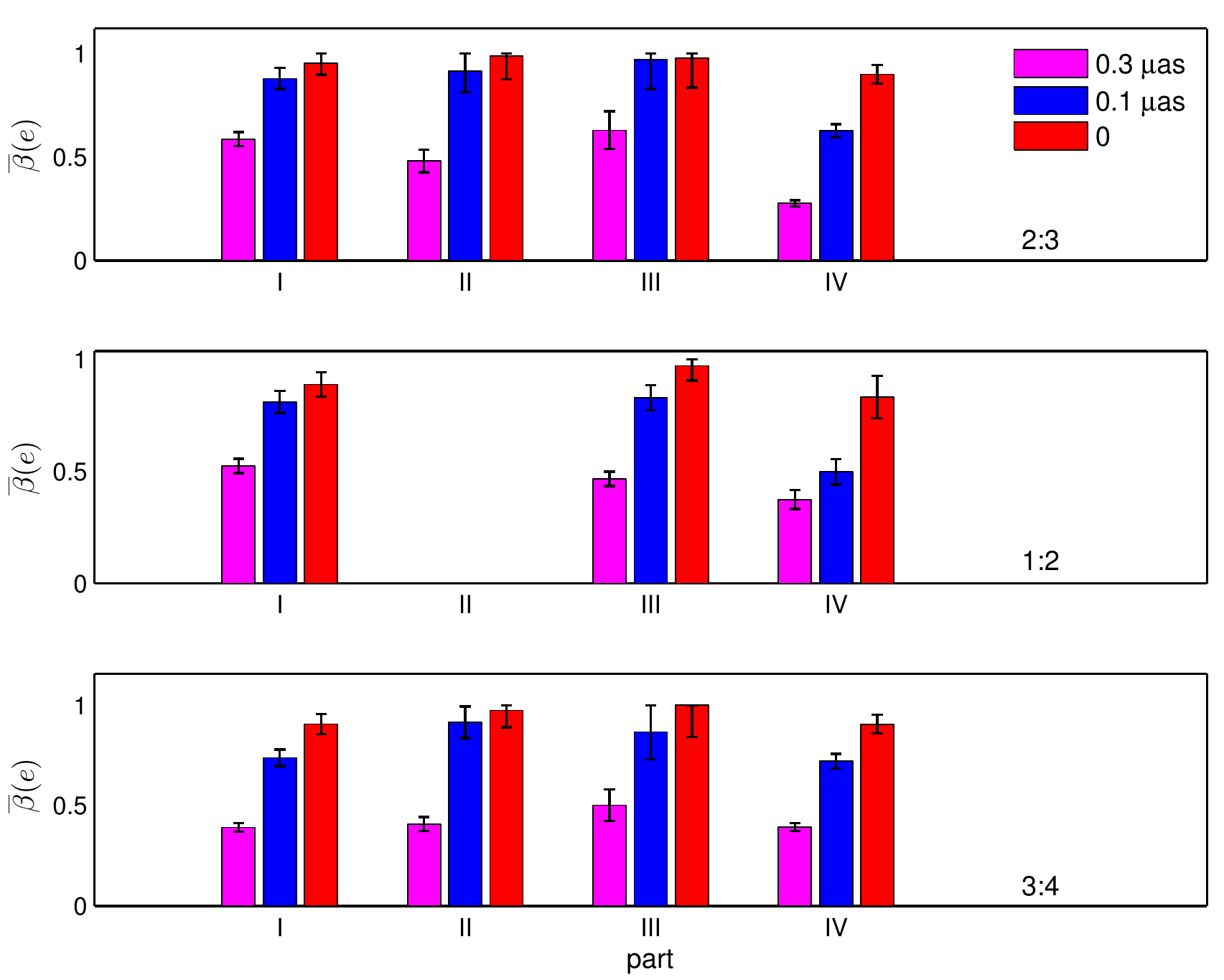}
\vspace{0cm}
\captionsetup{font={footnotesize}}
\caption{ Relations between the average resonance-reconstruction probabilities $\overline{\beta}(e)$  and eccentricities of the super-Earth pairs in MMR. The top, middle and the lower bar graphs are results of the 2:3, 1:2 and 3:4 MMRs, respectively. The red, blue and magenta colors represent observational errors $=$0, 0.1 $\mu$as, 0.3 $\mu$as, respectively. I,II,III and IV represent different ranges of eccentricities, i.e., I: $e_{1}>0.1$, $e_{2}>0.1$, II: $e_{1}<0.1$, $e_{2}>0.1$, III: $e_{1}>0.1$, $e_{2}<0.1$, IV: $e_{1}<0.1$, $e_{2}<0.1$. The number of samples in each part is larger than 20. The error bars displayed on the bins represent only the uncertainty due to Poisson statistics. For Super-Earth pairs in 1:2 MMR, there are few samples in part II, so we didn't show the results here.  \label{fig9} }.
\end{figure}

In addition to $\Delta$, eccentricities also have important influence on the probability of reconstructing a planet pair in MMR. We divide the systems in each MMR into 4 parts according to the eccentricities of the planets : I: $e_{1}>0.1$, $e_{2}>0.1$, II: $e_{1}<0.1$, $e_{2}>0.1$, III: $e_{1}>0.1$, $e_{2}<0.1$, IV: $e_{1}<0.1$, $e_{2}<0.1$. We calculate the average values of $\beta$(hereafter $\overline{\beta}(e)$) in each part with different observational errors for the Jupiter pairs and super-Earth pairs, as shown in Figure \ref{fig8} and Figure \ref{fig9}. To better illustrate the relation between MMR-reconstruction probability and eccentricity, we only calculate $\overline{\beta}(e)$ in eccentricity bins with the number of planet pairs larger than 20. The uncertainties due to Poisson statistics are shown as error bars displayed on the eccentricity bins. Different colors represent different observational errors. Obviously, $\overline{\beta}(e)$ with large error decreases in all cases.

We can see that eccentricities are essential for variation of $\overline{\beta}(e)$ in different parts. For planet pairs in 2:3 and 1:2 MMRs, $\overline{\beta}(e)$ in part I are the larger than $\overline{\beta}(e)$ in part IV. The increase of $\overline{\beta}(e)$ from part IV to part I is obvious for Jupiter pairs in 2:3 and 1:2 MMR. Stability constrains are quite strong in Jupiter pairs in 3:4 MMR, few planet pairs remain in part III and IV, $\overline{\beta}(e)$ in part I are larger than those in part II. For super-Earth pairs in 3:4 MMR, the increase of $\overline{\beta}(e)$ from part IV to part I is not obvious, this is because the average amplitudes of resonance angles are not well-distributed from part I to part IV, of which the influence on $\beta$ will be discussed in the following section. However, $\overline{\beta}(e)$ in part IV is still the smallest in 3:4 MMR. The positive correlation between $\overline{\beta}(e)$ and eccentricities indicates that eccentricities are important to MMR-reconstruction, because the precision of $\omega_{i}$ and $M_{i}$ are very sensitive to the precision of $e_{i}$(i=1,2). Simulations show that when eccentricities are smaller than 0.01, $\omega_{i,\rm fit}+M_{i,\rm fit}$ may deviate from the true value obviously. Even if $\omega_{i,\rm fit}+M_{i,\rm fit}$(i=1,2) equals to the true value, it's hard to decide both $\omega_{i}$ and $M_{i}$(i=1,2) accurately when $e_i<0.01$(i=1,2). The geometrical degeneracy of $\omega_{i}$ and $M_{i}$(i=1,2) makes us reconstruct planet pairs with small eccentricities in MMRs ambiguously. Large $e_{i}$ can avoid this degeneracy and result in more accurate $\omega_{i}$ and $M_{i}$. Besides, the resonance widths increase with eccentricities of the planet\citep{Deck2013}. With similar absolute fitting errors of eccentricities, planet pairs with large eccentricities are more probable to remain in MMRs. Therefore, $\overline{\beta}(e)$ increases with $e_{i}$(i=1,2).

\subsection{MMR-reconstruction with different resonance intensities}

\begin{figure*}[htbp]
\centering
\includegraphics[width=\textwidth]{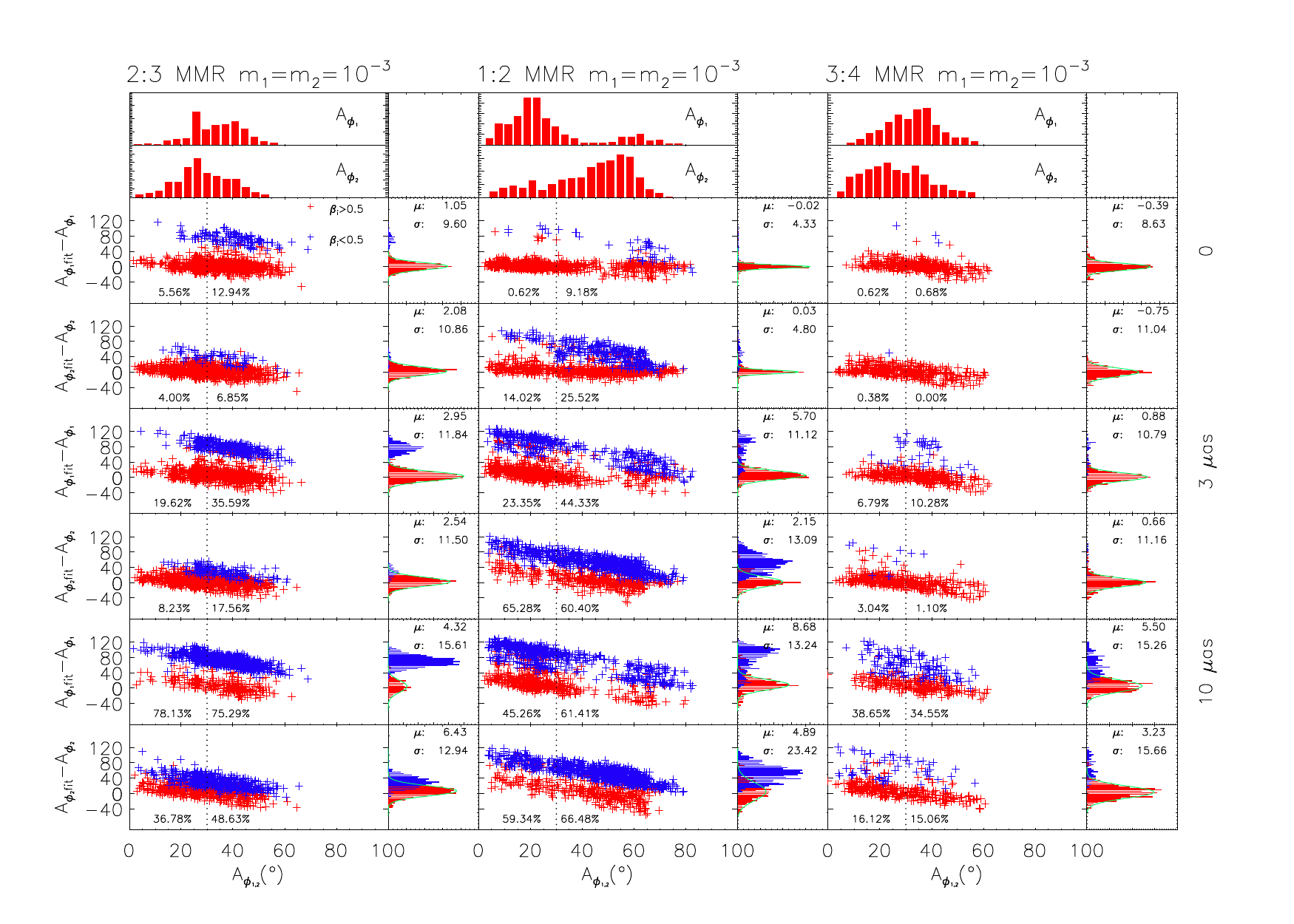}
\vspace{0cm}
\captionsetup{font={footnotesize}}
\caption{ The residuals of average amplitudes of resonance angles $\phi_1$ and $\phi_2$ in fitted systems($A_{\phi_{i}{\rm fit}}$) with those in real systems ($A_{\phi_{i}}$) of the Jupiter pairs. The top histograms are the distributions of the average amplitudes of resonance angles $\phi_1$ and $\phi_2$, i.e., $A_{\phi_{1}}$ and $A_{\phi_{2}}$.  The red crosses represent Jupiter pairs with $\beta_{i}>0.5$ and the blue crosses represent Jupiter pairs with $\beta_{i}<0.5$(i=1,2). The dot lines represent $A_{\phi_{i}}=30^\circ$(i=1,2). The values on the left side of the dot lines represent the blue fractions of Jupiter pairs with $A_{\phi_{i}}<30^\circ$ while the values on the right side represent the blue fractions of Jupiter pairs with $A_{\phi_{i}}>30^\circ$(i=1,2). $\mu$ and $\sigma$ are the mean value and standard deviation of residuals of $A_{\phi_{1}}$ and $A_{\phi_{2}}$ with $\beta_{i}>0.5$(i=1,2). The left, middle and right panels show the residuals of the 2:3, 1:2 and 3:4 MMRs, respectively. Every two panels from top down show residuals of  $A_{\phi_{2}}$ and $A_{\phi_{2}}$ with observational errors 0, 3 $\mu$as, 10 $\mu$as. \label{fig10} }.
\end{figure*}

\begin{figure*}[htbp]
\centering
\includegraphics[width=\textwidth]{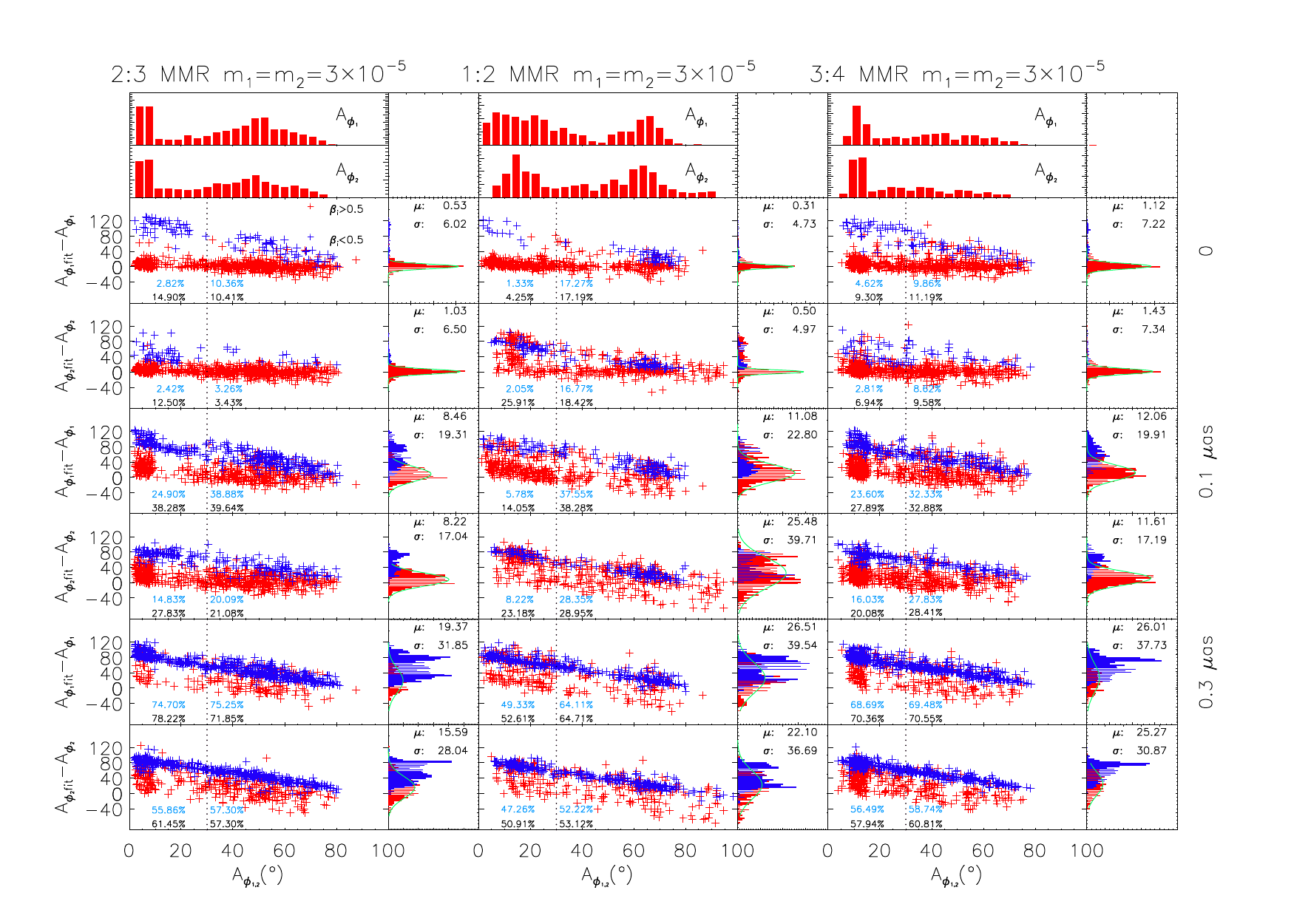}
\vspace{0cm}
\captionsetup{font={footnotesize}}
\caption{ Similar to Figure \ref{fig10} but for the super-Earth pairs. Besides, the values with blue color represent the blue fractions of planet pairs with $e_1>0.01$ and $e_2>0.01$. Every two panels from top down show residuals of the resonance angle $\phi_{1}$ and $\phi_{2}$ with observational errors 0, 0.1 $\mu$as, 0.3 $\mu$as. \label{fig11} }.
\end{figure*}

To check how well we reconstruct the MMRs, we compare the average amplitudes of $\phi_1$ and $\phi_2$ in fitted systems(hereafter $A_{\phi_{i},{\rm fit}}$) with those in real systems(hereafter $A_{\phi_{i}}$(i=1,2) during $2\times10^{4}$ years. In Figure \ref{fig10} and \ref{fig11}, the red crosses represent planet pairs with $\beta_i>0.5$ while the blue ones represent those with $\beta_i<0.5$(i=1,2). We adopt the Gaussian distribution to fit the residuals with $\beta_i>0.5$(i=1,2) and obtain mean value $\mu$ and standard deviation $\sigma$ for each kind of MMR at different observational errors. As shown in the Figure \ref{fig10} and \ref{fig11}, we find that with the increase of observational errors, both mean values and standard deviations becomes larger and larger, indicating that fitted resonance angles deviate more and more from the true values. Besides, the samples with $\beta_i<0.5$(i=1,2) have residuals far away from 0, which is reasonable, because in systems with $\beta_i<0.5$, the resonance angles $\phi_{i}$ only librate in $2\times10^{4}\cdot \beta_i$ years, the average amplitudes should be larger than those with $\beta_i>0.5$(i=1,2).

The blue crosses mostly come from real systems with large $A_{\phi_{i}}$(i=1,2) which means weak MMRs. To show the distribution of blue crosses clearly, we plot a dotted line in Figure \ref{fig10} and \ref{fig11}, which represents $A_{\phi_{i}}=30^{\circ}$(i=1,2), to divide the total samples into two categories. The values on the left and right side of the dotted line represent the fractions of the blue crosses in the two categories with $A_{\phi_{i}}<30^{\circ}$ and $A_{\phi_{i}}>30^{\circ}$(i=1,2). We find that the blue fractions on the right side are generally larger than those on the left side for Jupiter pairs in 2:3 and 1:2 MMRs. For Jupiter pairs in 3:4 MMR, blues fractions on each side are close because the stability will exclude part of systems with large $A_{\phi_{i}}$. For super-Earth pairs, there are much more systems with eccentricities $<$ 0.01 than systems containing Jupiter pairs, because planets with larger masses are more easily to excite their eccentricities in our sample simulations. According to section 4.3, small eccentricities lead to a small MMR-reconstruction probability, thus, the blue fraction with large $A_{\phi_{i}}$ is not always larger than the blue fraction with small $A_{\phi_{i}}$. To exclude the non-uniform distribution of systems with small eccentricities in the two categories with different $A_{\phi_{i}}$, we choose samples of the super-Earth pairs with $e_{1}>0.01$ and $e_{2}>0.01$ to recalculate the blue fractions of the two categories, which are shown as blue values in Figure \ref{fig11}. We find that the blue fractions for $A_{\phi_{i}}>30^{\circ}$ are larger than those for $A_{\phi_{i}}<30^{\circ}$(i=1,2) in 1:2, 2:3 and 3:4 MMRs.

Beyond that, there are obvious differences between the blue fractions of $A_{\phi_{1}}$ and $A_{\phi_{2}}$ for Jupiter pairs in 2:3 and 1:2 MMRs. Compared to $A_{\phi_{1}}$, $A_{\phi_{2}}$ concentrates upon a smaller value for 2:3 MMR. Naturally, with the same level of observational errors and similar deviation from the original values of $\phi_{1}$ and $\phi_{2}$, $\phi_{2}$ is more likely to remain in libration than $\phi_{1}$, therefore, the planet pairs with $\beta_2<0.5$ are less than those with $\beta_1<0.5$. On the contrary, for 1:2 MMR, $A_{\phi_{1}}$ concentrates upon a much smaller value than $A_{\phi_{2}}$, so $\phi_{1}$ is much easier to be reconstructed in libration than $\phi_{2}$. The positive correlation between the blue fraction and $A_{\phi_{i}}$(i=1,2) indicates that the stronger the intensities of the MMRs, the easier the MMRs can be reconstructed.\\

Analyses above indicate that the MMR-reconstruction probabilities are related with eccentricities and resonance intensities of the planet pairs. To better compare the difference of MMR-reconstruction probabilities between Jupiter pairs and super-Earth pairs, we calculate the fraction of planet pairs with $\beta>0.5$ among samples with appropriate eccentricities and strong intensities, i.e., eccentricities of both planets are larger than 0.01 and the average amplitudes of at least one resonance angle is smaller than $30^\circ$. The results are shown in Table \ref{tab3}. Except for planet pairs in 1:2 MMR with SNR=10, Jupiter pairs can be better reconstructed in MMRs than super-Earth pairs, especially when SNR=3, because of the Hill stability and planet pairs with larger masses have larger resonance width according to \citet{Deck2013}. In fact, it is quite hard to explain all the difference between super-Earth pairs and Jupiter pairs. Although we have confined the eccentricity and resonance intensity to compare the difference of MMR-reconstruction between Jupiter pairs and super-Earth pairs, we can not eliminate the sample bias between Jupiter and super-Earth pairs totally. A more refined sample control should be helpful to eliminate the exception.

\begin{table*}\footnotesize
\begin{center}
\begin{threeparttable}[b]
\captionsetup{font={footnotesize}}
\caption{fraction of stable planet pairs with $\beta>0.5$ among planet pairs with $e_i>0.01$(i=1,2) and $A_{\phi_1}(or A_{\phi_2})<30^\circ$}
\label{tab3}
\begin{tabular}{cccc|ccc}
\tableline\tableline\

&\multicolumn{3}{c|}{Jupiter pairs}&\multicolumn{3}{c}{super-Earth pairs}\\\hline
 observational errors &0&3 $\mu$as& 10 $\mu$as & 0 &0.1 $\mu$as &0.3 $\mu$as \\\hline
2:3&$96\%\pm1\%$&$92\%\pm2\%$&$63\%\pm1\%$&$97\%\pm2\%$&$85\%\pm1\%$&$44\%\pm3\%$\\

1:2&$99\%\pm1\%$&$79\%\pm2\%$&$64\%\pm4\%$&$99\%\pm1\%$&$97\%\pm3\%$&$55\%\pm6\%$\\

3:4&$100\%\pm1\%$&$97\%\pm1\%$&$84\%\pm1\%$&$97\%\pm1\%$&$84\%\pm1\%$&$44\%\pm1\%$\\\hline

\end{tabular}
  \end{threeparttable}
  \begin{flushleft}
Note. The uncertainties are calculated similar to those in Table \ref{tab1}.
\end{flushleft}
\end{center}
\end{table*}

\section{The false alarm probability of planet pairs in or near MMRs analysis}

In section 4, we have investigated the probability to reconstruct planet pairs in MMRs. Consequently, the fraction of well reconstructed planet pairs with $\beta>0.5$ in MMRs(hereafter denoted as $P_{1-1}$) can be obtained according to previous section. Actually, some systems are not in but near MMRs. We are also interested in the FAPs (hereafter denoted as $P_{0-1}$) of mistaking planet pairs near MMRs for planet pairs in MMRs.

To obtain the FAPs of mistaking near MMR systems for systems in MMR, we simulate 1600 Jupiter pairs with $\Delta<0.04$, and 1600 super-Earth pairs with $\Delta<0.02$ near each kind of MMR. All samples near MMRs adopted to estimate the FAPs($P_{0-1}$) in our simulations are stable and not in MMRs in $2\times10^{4}$ years.  In our simulations, $P_{0-1}$ is calculated as the fraction of planet pairs which are fitted to be in MMRs with a probability $\beta>$ 0.5. $P_{0-1}$ for different MMRs at different ranges of $\Delta_{\rm fit}$(or $\Delta$) are shown in Table \ref{tab4}(Jupiter pairs) and Table \ref{tab5}(super-Earth pairs). $\Delta_{\rm fit}$ is calculated with the fitted average periods of the planets in $2\times10^4$ years. Almost all $\Delta_{\rm fit}$ of Jupiter pairs are smaller than 0.04, and  $\Delta_{\rm fit}$ of super-Earth pairs are smaller than 0.02. We do not analysis $P_{0-1}$ for Jupiter pairs near the 3:4 MMR because of their weak stabilities, i.e., $P_{0-1}\sim0$ when considering the long time stabilities of these systems. For Jupiter pairs and super-Earth pairs in each MMR, we divide the samples into three ranges according to $\Delta_{\rm fit}$ or $\Delta$, and calculate $P_{0-1}$ in each range of $\Delta_{\rm fit}$(outside of the brackets) and $\Delta$(in the brackets). $P_{0-1}$ for all samples are shown in the fourth row of each MMR in Table \ref{tab4} and Table \ref{tab5}.

\begin{table*}[htbp]
\footnotesize
\begin{center}
\begin{threeparttable}[b]
\captionsetup{font={footnotesize}}
\caption{$P_{0-1}$ of the two Jupiter system\tnote{1} \label{tab4}}
\begin{tabular}{cccccc}
\tableline\tableline
\multicolumn{2}{c}{} &  \multicolumn{3}{c}{observational error} \\ \hline
\multicolumn{2}{c}{$\Delta_{fit}$($\Delta$)\tnote{2}} &\bf{0}  & 3 $\mu$as&{10 $\mu$as} \\\hline
2:3&{$<0.01$}&$5\%\pm1\%$($3\%\pm1\%$) &$36\%\pm1\%$($30\%\pm1\%$)&$42\%\pm1\%$($46\%\pm1\%$)\\
 &$0.01\thicksim0.02$&$3\%\pm1\%$($3\%\pm1\%$) &$10\%\pm2\%$($25\%\pm1\%$)&$14\%\pm1\%$($33\%\pm1\%$)\\
&$0.02\thicksim0.04$ &$0.3\%\pm1\%$($5\%\pm2\%$)&$8\%\pm2\%$($18\%\pm1\%$) &$0\%\pm1\%$($38\%\pm1\%$)\\
&$\le0.04$&$3\%\pm1\%$($3\%\pm1\%$)&$27\%\pm1\%$($26\%\pm1\%$)&$38\%\pm1\%$($38\%\pm1\%$)\\\hline
1:2 &$<5\times10^{-3}$&$37\%\pm2\%$($12\%\pm2\%$)&$44\%\pm4\%$($8\%\pm1\%$)&$51\%\pm2\%$($19\%\pm2\%$)\\
   &$5\times10^{-3}\thicksim0.01$&$18\%\pm1\%$($15\%\pm2\%$)&$13\%\pm2\%$($19\%\pm1\%$)&$10\%\pm1\%$($35\%\pm1\%$)\\
   &$0.01\thicksim0.04$&$2\%\pm1\%$($12\%\pm1\%$)&$4\%\pm1\%$($16\%\pm1\%$)&$0\%\pm1\%$($32\%\pm3\%$)\\
&$\le0.04$&$14\%\pm1\%$($13\%\pm1\%$)&$17\%\pm1\%$($16\%\pm1\%$)&$33\%\pm1\%$($32\%\pm1\%$)\\\hline

\end{tabular}
\begin{tablenotes}
    \item [1] $P_{0-1}$ is the possibility of mistaking a planet system near MMR for one in MMR.
    The values outside of the brackets are $P_{0-1}$s calculated via $\Delta_{\rm fit}$ and values inside of the brackets are $P_{0-1}$s calculated via $\Delta$. $P_{0-1}$ is different from $\rm \mathscr{F}_{0-1}$ in Table \ref{tab7} and Table \ref{tab8}.
	\item [2] $\Delta_{\rm fit}$ is the normalized distance to resonance center of the fitted planet pairs, $\Delta_{\rm fit}=(j-1)P_{\rm fit,2}/(jP_{\rm fit,1})-1$
   \end{tablenotes}
  \end{threeparttable}
  \begin{flushleft}
Note. The uncertainties are calculated similar to those in Table \ref{tab1}.
\end{flushleft}
\end{center}
\end{table*}

\begin{table*}[htbp]
\footnotesize
\begin{center}
\begin{threeparttable}[c]
\captionsetup{font={footnotesize}}
\caption{$P_{0-1}$ of the two super-Earth system\tnote{1}. \label{tab5}}
\begin{tabular}{ccccc}
\tableline\tableline
\multicolumn{2}{c}{} &  \multicolumn{3}{c}{observational error}\\ \hline
\multicolumn{2}{c}{$\Delta_{fit}(\Delta)$} &{0}  & 0.1 $\mu$as&0.3 $\mu$as \\\hline
2:3&$<5\times10^{-4}$&$6\%\pm1\%$($5\%\pm2\%$)&$24\%\pm1\%$($14\%\pm1\%$)&$45\%\pm1\%$($17\%\pm3\%$)\\
 &$5\times10^{-4}\thicksim5\times10^{-3}$&$2\%\pm1\%$($1\%\pm1\%$)&$3\%\pm1\%$($16\%\pm1\%$)&$3\%\pm1\%$($17\%\pm1\%$)\\
&$5\times10^{-3}\thicksim0.02$ &$0.5\%\pm1\%$($1\%\pm1\%$)&$0.00\%\pm1\%$($2\%\pm1\%$)&$0.0\%\pm1\%$($13\%\pm1\%$)\\
&$\le0.02$&$3\%\pm1\%$($2\%\pm1\%$)&$9\%\pm1\%$($8\%\pm1\%$)&$21\%\pm1\%$($15\%\pm1\%$)\\\hline
1:2 &$<1.6\times10^{-4}$&$34\%\pm2\%$($17\%\pm4\%$)&$60\%\pm1\%$($26\%\pm1\%$)&$68\%\pm1\%$($32\%\pm4\%$)\\
   &$1.6\times10^{-4}\thicksim5\times10^{-3}$&$20\%\pm2\%$($25\%\pm3\%$)&$24\%\pm1\%$($33\%\pm1\%$)&$13\%\pm1\%$($26\%\pm2\%$)\\
   &$5\times10^{-3}\thicksim0.02$&$1\%\pm1\%$($2\%\pm1\%$)&$1\%\pm1\%$($12\%\pm2\%$)&$0.1\%\pm1\%$($19\%\pm1\%$)\\
&$\le0.02$&$14\%\pm1\%$($13\%\pm1\%$)&$23\%\pm2\%$($22\%\pm1\%$)&$31\%\pm1\%$($24\%\pm1\%$)\\\hline

3:4 &$<5\times10^{-4}$&$7\%\pm2\%$($5\%\pm1\%$)&$29\%\pm1\%$($17\%\pm2\%$)&$45\%\pm2\%$($20\%\pm1\%$)\\
&$5\times10^{-4}\thicksim5\times10^{-3}$&$3\%\pm1\%$($6\%\pm1\%$)&$1\%\pm1\%$($10\%\pm2\%$)&$1\%\pm2\%$($10\%\pm4\%$)\\
&$5\times10^{-3}\thicksim0.02$&$0.7\%\pm1\%$($1\%\pm1\%$) &$0.6\%\pm1\%$($6\%\pm1\%$)&$0.4\%\pm1\%$($13\%\pm1\%$)\\
&$\le0.02$&$4\%\pm1\%$($3\%\pm1\%$)&$12\%\pm1\%$($10\%\pm1\%$)&$23\%\pm2\%$($15\%\pm1\%$)\\\hline

\end{tabular}
\begin{tablenotes}
    \item [1] The same with Table \ref{tab4}.

   \end{tablenotes}
  \end{threeparttable}
  \begin{flushleft}
Note. The uncertainties are calculated similar to those in Table \ref{tab1}.
\end{flushleft}
\end{center}
\end{table*}

The fourth rows of each MMR in Table \ref{tab4} and Table \ref{tab5} show that there is a positive correlation between $P_{0-1}$ and observational error when $P_{0-1}$ is calculated via $\Delta$. As the larger the observational error is, the further the fitted orbital parameters deviate from their true values, thus the fitted planet pairs can arrive some islands of MMRs far away from the initial position in phase space and they are probably in MMRs.
Unlike the positive correlation between $P_{0-1}$ and observational error, there is a negative correlation between $P_{0-1}$ and $\Delta_{\rm fit}$, i.e., the larger the $\Delta_{\rm fit}$ is, the further planet pair is away from the MMR center, thus it's less likely to be mistaken for a planet pair in MMR. However, $P_{0-1}$ have no obvious correlation with $\Delta$. Because $\Delta_{\rm fit}$ in false alarm cases are usually small, while $\Delta$ are widely distributed. Only few cases with large $\Delta_{\rm fit}$ are mistaken for systems in MMR. When $\Delta_{\rm fit}$ is large enough, i.e. $\Delta_{\rm fit}>0.02$ for Jupiter pairs, $P_{0-1}$ decrease to smaller than $10\%$, and $\Delta_{\rm fit}>0.005$ for super-Earth pairs, $P_{0-1}$ decrease to about $1\%$.

When we detect a planet pair with period ratio near 1:2, 2:3 or 3:4 MMR, and simulation shows that it is in MMR based on the fitted parameters, the detected planet pair in MMR might be a false alarm. To calculate the FAP $\rm \mathscr{F}_{0-1}$ for a detected planet system in MMR, we need the values of both $P_{0-1}$ and $P_{1-1}$. If we assume the same Number $N_p$ of planet pairs in or near MMRs, $N_p \cdot P_{0-1}$ planet pairs near MMRs will be mistaken as planet pairs in MMRs, while $N_p \cdot P_{1-1}$ planet pairs in MMRs can be well reconstructed. Finally, $\rm \mathscr{F}_{0-1}$ is expressed as:
\beq
\rm \mathscr{F}_{0-1}=\frac{P_{0-1}}{P_{0-1}+P_{1-1}}.
 \label{fap1}
 \eeq
Note that the meaning of $P_{0-1}$ is different with that of $\rm \mathscr{F}_{0-1}$. From Equation (\ref{fap1}), we can see that even if $P_{1-1}=1$, $\rm \mathscr{F}_{0-1}$ is greater than 0, but smaller than $P_{0-1}$.

On the other hand, there is another FAP when we reconstruct a planet system near MMR. Take $P_{0-0}$ as the probability of reconstructing a system near but not in MMR, and take $P_{1-0}$ as the probability of mistaking an in MMR system for a near MMR system. Similar with the derivation of $\rm \mathscr{F}_{0-1}$, the FAP for a near MMR system $\rm \mathscr{F}_{1-0}$ is expressed as:
\beq
\rm \mathscr{F}_{1-0}=\frac{P_{1-0}}{P_{0-0}+P_{1-0}}.
 \label{fap2}
 \eeq
It's easy to obtain $P_{1-0}=1-P_{1-1}$ and $P_{0-0}=1-P_{0-1}$. In observations, only $\Delta_{\rm fit}$ can be obtained, so it's suitable to adopt the values of $P_{1-1}$, $P_{1-0}$, $P_{0-0}$ and $P_{0-1}$ calculated via $\Delta_{\rm fit}$. $P_{1-1}$ in Table \ref{tab6} are slightly larger than values in the last column in Table \ref{tab1} and Table \ref{tab2}, because they are calculated among planet pairs with $\Delta_{\rm fit}$ in the same range of $\Delta_{\rm fit}$ shown in the third rows in Table \ref{tab4} and Table \ref{tab5}.

\begin{table*}[htbp]
\begin{center}
\begin{threeparttable}[c]
\captionsetup{font={footnotesize}}
\caption{$P_{1-1}$ of different MMRs at different observational errors.}

\label{tab6}
\begin{tabular}{ccccc|cccc}
\tableline\tableline
&\multicolumn{4}{c|}{Jupiter pairs}&\multicolumn{4}{c}{super-Earth pairs}\\\hline
observational errors&&{0}&{3 $\mu$as}&10 $\mu$as&&{0}&0.1 $\mu$as&0.3 $\mu$as\\\hline
2:3 MMR&$\Delta_{fit}<0.01$&$99\%\pm1\%$&$96\%\pm1\%$&$62\%\pm1\%$&$\Delta_{fit}<5\times10^{-4}$&$98\%\pm1\%$&$87\%\pm2\%$&$56\%\pm1\%$\\
1:2 MMR&$\Delta_{fit}<5\times10^{-3}$&$99\%\pm1\%$&$81\%\pm4\%$&$69\%\pm1\%$&$\Delta_{fit}<1.6\times10^{-4}$&$95\%\pm1\%$&$92\%\pm1\%$&$66\%\pm5\%$\\
3:4 MMR&$\Delta_{fit}<1\times10^{-3}$&$100\%\pm1\%$&$97\%\pm1\%$&$87\%\pm4\%$&$\Delta_{fit}<5\times10^{-4}$&$96\%\pm1\%$&$86\%\pm1\%$&$51\%\pm5\%$\\\hline
\end{tabular}
\footnotesize

  \end{threeparttable}
\begin{flushleft}
Note. The uncertainties are calculated similar to those in Table \ref{tab1}.
\end{flushleft}
\end{center}
\end{table*}

Table \ref{tab7} and Table \ref{tab8} show the final $\rm \mathscr{F}_{0-1}$ and $\rm \mathscr{F}_{1-0}$ of a planet system detected in or near MMRs. Generally, the larger the observational errors are, the larger $\rm \mathscr{F}_{0-1}$ and $\rm \mathscr{F}_{1-0}$ are. For both Jupiter and super-Earth pairs, $\rm \mathscr{F}_{0-1}$ and $\rm \mathscr{F}_{1-0}$ are sensitive to the observational errors. $\rm \mathscr{F}_{0-1}$ of Jupiter pairs and super-Earth pairs in 1:2 MMR are very similar,  which are larger than $20\%$ even without observational errors. With the same observational errors, $\rm \mathscr{F}_{0-1}$ and $\rm \mathscr{F}_{1-0}$ for planet pairs in 2:3 and 3:4 MMRs are smaller than those in 1:2 MMR. Note that the particularity of large FAP for planet pairs in 1:2 MMR is mainly induced by the significant large $P_{0-1}$. As we have mentioned before, the 1:2 period ratio makes it hard to fit planet pairs as well as planet pairs with other period ratio. So it's likely to mistake planet pairs near 1:2 MMR for those in 1:2 MMR. When SNR $\sim$ 3, both $\rm \mathscr{F}_{0-1}$ and $\rm \mathscr{F}_{1-0}$ are larger than $30\%$, therefore, if we detect a planet system in or near MMR with low SNR, the system should be checked carefully.

\begin{table*}[htbp]
\begin{center}
\begin{threeparttable}[c]
\begin{tabular}{ccc|cc}
\tableline\tableline
&\multicolumn{2}{c|}{$\rm \mathscr{F}_{0-1}$\tnote{1}}&\multicolumn{2}{c}{$\rm \mathscr{F}_{1-0}$\tnote{2}}\\\hline
observational error & 2:3($\Delta_{fit}<0.01$) & 1:2 ($\Delta_{fit}<5\times10^{-3}$)& 2:3($\Delta_{fit}<0.01$) & 1:2 ($\Delta_{fit}<5\times10^{-3}$)\\\hline
without error &$5\%\pm1\%$ &$27\%\pm1\%$&$1\%\pm1\%$&$1\%\pm2\%$ \\
3 $\mu$as &$27\%\pm1\%$ &$35\%\pm1\%$&$6\%\pm1\%$&$25\%\pm3\%$  \\
10 $\mu$as&$40\%\pm1\%$ &$42\%\pm1\%$ & $39\%\pm1\%$&$38\%\pm1\%$\\\hline

\end{tabular}
\footnotesize
\begin{tablenotes}
    \item [1] $\rm \mathscr{F}_{0-1}$ is the false alarm probability when we detect a planet pair in MMR. It is calculated on the\\ basis of the possibility we mistake a planet pair near but not in MMR for the one in MMR and \\the possibility we detect a true planet pair in MMR.
    \item [2] $\rm \mathscr{F}_{1-0}$ is the false alarm probability when we detect a planet pair near but not in MMR. It is \\calculated on the basis of the possibility we mistake a planet pair in MMR for the one near MMR and \\the possibility we detect a true planet pair near but not in MMR.
   \end{tablenotes}
  \end{threeparttable}
  \begin{flushleft}
Note. The uncertainties are according to uncertainties in Table \ref{tab4} and Table \ref{tab6}. As we have $\rm \mathscr{F}_{0-1}=\frac{P_{0-1}}{P_{0-1}+P_{1-1}}$, the uncertainty of $\rm \mathscr{F}_{0-1}$ can be estimated as $\frac{P_{1-1}dP_{0-1}-P_{0-1}dP_{1-1}}{(P_{0-1}+P_{1-1})^2}$.
Similarly, $\rm \mathscr{F}_{1-0}=\frac{P_{1-0}}{P_{1-0}+P_{0-0}}$, the uncertainty of $\rm \mathscr{F}_{1-0}$ is $\frac{P_{0-0}dP_{1-0}-P_{1-0}dP_{0-0}}{(P_{1-0}+P_{0-0})^2}$.
\end{flushleft}
\end{center}
\captionsetup{font={footnotesize}}
\caption{FAP of the two Jupiter system \label{tab7}}
\end{table*}

\begin{table*}[htbp]
\begin{center}
\begin{threeparttable}[c]
\tiny
\begin{tabular}{cccc|ccc}
\tableline\tableline
&\multicolumn{3}{c|}{$\rm \mathscr{F}_{0-1}$\tnote{1}}&\multicolumn{3}{c}{$\rm \mathscr{F}_{1-0}$\tnote{2}}\\\hline
observational error  & 2:3($\Delta_{fit}<5\times10^{-4}$) & 1:2 ($\Delta_{fit}<1.6\times10^{-4}$) & 3:4 ($\Delta_{fit}<5\times10^{-4}$)& 2:3($\Delta_{fit}<5\times10^{-4}$) & 1:2 ($\Delta_{fit}<1.6\times10^{-4}$) & 3:4 ($\Delta_{fit}<5\times10^{-4}$)\\\hline
without error&$6\%\pm2\%$ &$26\%\pm1\%$ &$7\%\pm1\%$ & $2\%\pm1\%$ &$7\%\pm1\%$ & $4\%\pm1\%$ \\
0.1 $\mu$as  &$21\%\pm1\%$ &$39\%\pm1\%$ &$25\%\pm1\%$ & $14\%\pm2\%$ & $16\%\pm1\%$ & $16\%\pm1\%$\\
0.3 $\mu$as &$44\%\pm1\%$  &$50\%\pm2\%$  &$47\%\pm1\%$ & $44\%\pm1\%$ & $52\%\pm3\%$ & $47\%\pm2\%$\\\hline

\end{tabular}
\tiny
\begin{tablenotes}
    \item [1] The same with $\rm \mathscr{F}_{0-1}$ in Table \ref{tab7}.
    \item [2] The same with $\rm \mathscr{F}_{1-0}$ in Table \ref{tab7}.
   \end{tablenotes}
  \end{threeparttable}
  \begin{flushleft}
Note. The uncertainties are according to uncertainties in Table \ref{tab7}.
\end{flushleft}
\end{center}
\captionsetup{font={footnotesize}}
\caption{FAP of the two super-Earth system \label{tab8}}
\end{table*}

\section{The potential of discovering planet pairs in MMRs}

After calculating the MMR-reconstruction probabilities, we can estimate the number of planet pairs in MMRs($N_{\rm MMR}$) which can be measured by astrometry if we know the frequency of Jupiter pairs and super-Earth pairs in MMRs around nearby stars.

Based on observations before Kepler Mission, \citet{Casertano2008} estimate the number of multiple planet systems that GAIA can detect. In their paper, they list all the multiple planet systems detected and calculate the fraction of the multiple planet systems which meet the condition SNR $>3$ with single-measurement precision set to be 8 $\mu$as. Then they extrapolate the results to the planet systems GAIA can detected and finally estimate the number of multiple planet systems they can find. However, in this paper, it's hard to estimate $N_{\rm MMR}$ in the same way due to lack of samples with parallax measurements. Among 415 \footnote{exoplanet.org} multiple planet systems detected, 76 of them have parallax measurements, and only 27 systems have planet pairs near MMRs.
The samples are very rare and no super-Earth pairs near MMRs appear in the 27 systems, so we choose another way to estimate $N_{\rm MMR}$.

 The number of planet pairs in MMR reconstructed by astrometry measurements can be expressed as: $N_{\rm MMR}=N_{\star}\times f_1\times f_2\times f_3\times f_4\times f_5$. $N_{\star}$ is the number of target stars, here we adopt $N_{\star}$=$3\times10^4$ based on the fact that there are more than $3\times10^4$ bright stars(V $<$ 10) within 30 pc (The Hipparcos and Tycho Catalogues). $f_1$ is the probability that a star host planets. $f_2$ is the probability that the planets are in multiple planet systems. $f_3$ is the probability that there are planets in MMRs in multiple planet systems. $f_4$ is the probability of planets in MMRs with Jupiter-like or super-Earth-like masses. $f_5$ is the probability that the planets in MMRs can be reconstructed by astrometry.

According to \citet{Cassan2012}, each Milky Way star hosts at least one planet, i.e., $f_1$ is set as $100\%$. We calculate $f_2, f_3$, and $f_4$ based on the planets discovered so far. According to observations of the Kepler mission, $f_2\sim41\%$\footnote{exoplanet data used in this section are from exoplanets.org,}. There is observational bias in Kepler mission which tends to discover planets close enough to the host star, the planets far away from the host star have smaller probability to be detected. Therefore, planet pairs in or near MMRs in observation mostly have semi-major axis $<$0.5 AU. For planets detected by transit, the occurrence rates of terrestrial planets decrease from 50 to 300 days, however, occurrence rates for planets with larger periods are hardly constrained \citep{Burke2015}. For planets detected by radial velocity, \citet{Cumming2008} found evidence for a sharp rise in occurrence of planets with periods $\ge1$ year. \citet{Winn2015} summarized the basic picture of planet probability density: giant planets have a probability density nearly constant in $logP$ between 2-2000 days, while smaller planets(1-4 $R_{\oplus}$) have a probability nearly constant in $logP$ between 10 and 300 days. Here, we simply assume that the occurrences of MMRs far away from the host star(1 AU) are similar to those near the host star. As few planets in MMRs have been confirmed, we set $f_3$ as the probability of near MMR planet pairs in multiple planet systems.  It is reasonable because many researches \citep{Lithwick2012,Batygin2013,Xie2014,Chatterjee2015} hint that planet pairs in MMRs can evolve into the observed MMR offset due to several mechanisms such as tidal dissipation and planet-planetesimal disk interaction. Currently, 415 \footnote{exoplanet.org} multiplanet systems are detected, and 135, 91 and 20 planet systems contain planet pairs near 2:3, 1:2 and 3:4 MMRs, i.e., $f_3=21.9\%, 32.5\%$ and $4.8\%$ for 2:3, 1:2 and 3:4 MMRs, respectively. Besides, among the planet pairs near MMRs, the fraction of both planets with masses $5 M_{\rm Earth} \sim 20 M_{\rm Earth}$ are $23.69\%$, the fraction of both planets with masses $0.5 M_{\rm J} \sim 2 M_{\rm J}$ are $2.79\%$. We choose $f_4=2.79\%$ for Jupiter pairs and $f_4=23.69\%$ for super-Earth pairs.

In our simulations, planet pairs in or near MMRs have inclinations between 0 and $10^\circ$, however, planet pairs in MMRs with inclination $\sim90^\circ$ can also be reconstructed with a certain probability. We do simulations for a super-Earth pair in 2:3 MMR with their inclinations increase from 0 to 90$^\circ$ and find that the MMR-reconstruction probability decrease if inclinations $\ge50^\circ$. Here we simply assume the MMR-reconstruction probability decrease linearly with increase of inclinations, i.e., $f_5(i)=f_5(i\simeq0^\circ)(1-|i|/(\pi/2))(i=[-\pi/2,\pi/2])$. $f_5(i\simeq0^\circ)$ is approximated by the MMR-reconstruction probability of planet pairs with nearly face-on orbits which has been calculated in Section 4, i.e., last columns in Table \ref{tab1} and \ref{tab2}. Besides, assuming a uniformly distribution of planet's orbital angular momentum vector, the probability density of inclination $dP(i)/di=\sin{|i|}/2(i=[-\pi/2,\pi/2])$. So we have $f_5=f_5(i\simeq0^\circ)\int^{\pi/2}_{-\pi/2}\sin{|i|}/2(1-|i|/(\pi/2))di\simeq0.36f_5(i\simeq0^\circ)$. Although $f_5(i\simeq0^\circ)$ are obtained by simulation of planet pairs near 1 AU, planet pairs at different locations will lead to the same results with the same SNR, if we rescale the observational errors and data samplings consistant with the locations of the inner planet.

Finally, we estimate the probabilities of discovering and reconstructing the planet systems by astrometry method, as shown in Table \ref{tab9}. As all the planet systems in our simulations are at 30 pc, the MMR-reconstruction probabilities are the inferior limits. The number of Jupiter pairs in MMRs can be detected and reconstructed by us are much less than that of the super-Earth pairs. It's reasonable, for planet systems containing two giants are rare in observations. With observational SNR$=3$, we can find tens of giant planet pairs in 2:3 and 1:2 MMRs. The reconstruction of super-Earth pairs in MMRs require higher precision to reach SNR$\sim$3, hundreds of super-Earth pairs in 2:3 or 1:2 MMRs will be identified, respectively. Planet pairs in 3:4 MMR are much less. With a higher SNR$=10$, about 1.2 times of Jupiter pairs and 1.8 times of super-Earth pairs in MMRs can be reconstructed than results with SNR$=3$. Jupiter pairs in 3:4 MMRs are strictly constrained by stability. Even with SNR$=3$, $f_5(i\simeq0^\circ)\sim100\%$, and all Jupiter pairs near 3:4 MMR can be detected by the direct orbital fitting or dynamical analysis, therefore, improvement of SNR can not enhance the number of Jupiter pairs. The discovery and reconstruction of planet pairs in MMRs are essential for planet formation and evolution theories. High precision of astrometry will lead us to make great processes.

\begin{table}
\begin{center}

\begin{threeparttable}[b]
\captionsetup{font={footnotesize}}
\caption{Number of planet pairs in MMRs can be\\ detect and reconstructed by astrometry in 30 pc.}

\label{tab9}
\begin{tabular}{ccc|cc}
\tableline\tableline
&\multicolumn{2}{c|}{Jupiter pairs}&\multicolumn{2}{c}{super-Earth pairs}\\\hline
SNR&$\sim$10&$\sim$3&$\sim$10&$\sim$3\\\hline
2:3 MMR&24&16&178&98\\
1:2 MMR&29&24&275&171\\
3:4 MMR&6&6&39&20\\\hline
\end{tabular}
\footnotesize

  \end{threeparttable}
\end{center}
\end{table}

\section{Even and uneven data cadence}

The simulations above are all carried out with even data cadence. In fact, most observations have uneven data cadence because of lots of realistic observational limits. To find out the influence of uneven data cadence on planet detection by astrometry method, we compare the differences between even and uneven data cadence in this section. Although the uneven data cadences discussed here are not realistic cadences schemes for Gaia and STEP, it's important to explore how large the influence is.

\begin{figure}
\centering
\includegraphics[width=\textwidth]{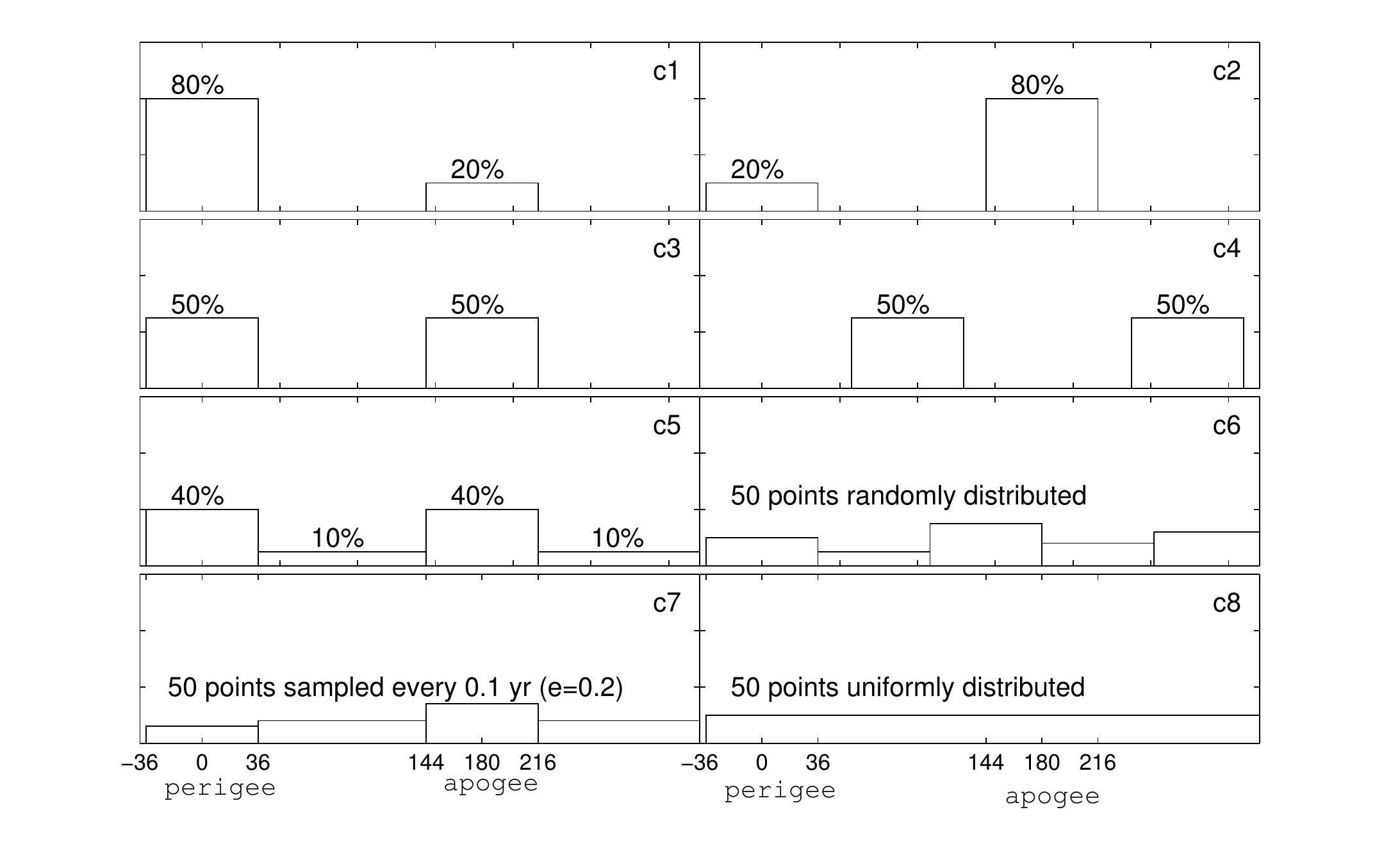}
\vspace{0cm}
\captionsetup{font={footnotesize}}
\caption{Diagrammatic sketch of the eight kinds of data cadence. c1-c8 represent data cadence c1-c8 described in section 7,respectively. \label{fig12} }.
\end{figure}

For single planet systems, we simulate 100 super-Earth systems and 100 Jupiter systems which are 30 pc from us. All the planets are 0.9 AU from the host star and their eccentricities are distributed from 0.01 to 0.5. All observations have a set of 50 data points.  We choose 8 different data cadences of simulated astrometry data as follows:
\begin{itemize}
\item[c1:] 80$\%$ data points are randomly distributed near the perigee, i.e., $-43.2^\circ<f<43.2^\circ$, where $f$ is the true anomaly (hereafter the same). 20$\%$ are randomly distributed near the apogee, i.e., $129.6^\circ<f<216^\circ$ (hereafter the same);
\item[c2:] 20$\%$ are randomly distributed near the perigee, while 80$\%$ are randomly distributed near the apogee;
\item[c3:] 50$\%$ are randomly distributed near the perigee, while 50$\%$ are near the apogee;
\item[c4:] 50$\%$ are randomly distributed near the mid point of the apogee and perigee, i.e., $43.2^\circ<f<129.6^\circ$. The others are randomly distributed on the opposite side, where $230.4^\circ<f<316.8^\circ$.
\item[c5:] 40$\%$ are randomly distributed near the perigee, while 40$\%$ are randomly distributed near the apogee. The other 20$\%$ are randomly distributed in the left regions;
\item[c6:] $f$ of all data points are randomly distributed;
\item[c7:] Times of all data points are uniformly distributed, i.e., even data cadence adopted before this section.
\item[c8:] All data points are distributed with uniform orbital phase coverage, i.e., there is one data point in each range of $f$ with a width of $7.2^\circ$;
\end{itemize}
The diagrammatic sketches of the 8 data cadences are shown in Figure \ref{fig12}.

\begin{figure}
\centering
\includegraphics[width=\textwidth]{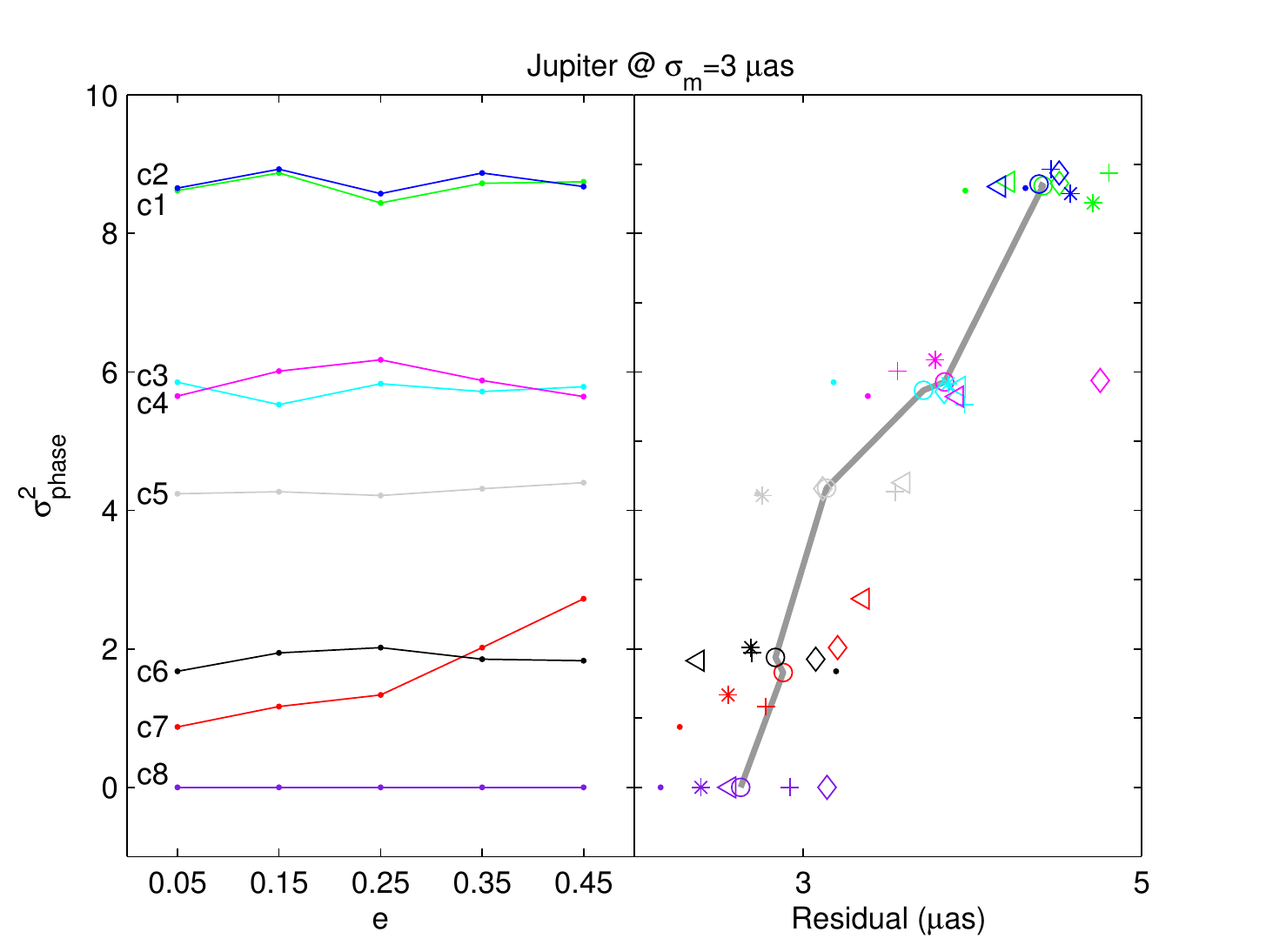}
\vspace{0cm}
\captionsetup{font={footnotesize}}
\caption{The left panel shows the variances of the data cadence at different eccentricities. The green, blue, cyan, magenta, light grey, dark, red and purple lines represent the data cadence c1-c8, respectively. The right panel shows the difference between the true and fitted astrometric signature caused by the Jupiter with a standard deviation of observational error $\sigma_m=3$ $\mu$as at different variances $\sigma^2_{\rm phase}$ and eccentricities $e$ of the data cadence. Different colors represent data cadences the same with those in the left panel. The symbols dot, cross, asterisk, diamond and left triangle represent the mean variance and residuals with mean eccentricities e=0.05, 0.15, 0.25, 0.35 and 0.45, respectively. The circles are the mean values with any eccentricities. \label{fig13} }.
\end{figure}

\begin{figure}
\centering
\includegraphics[width=\textwidth]{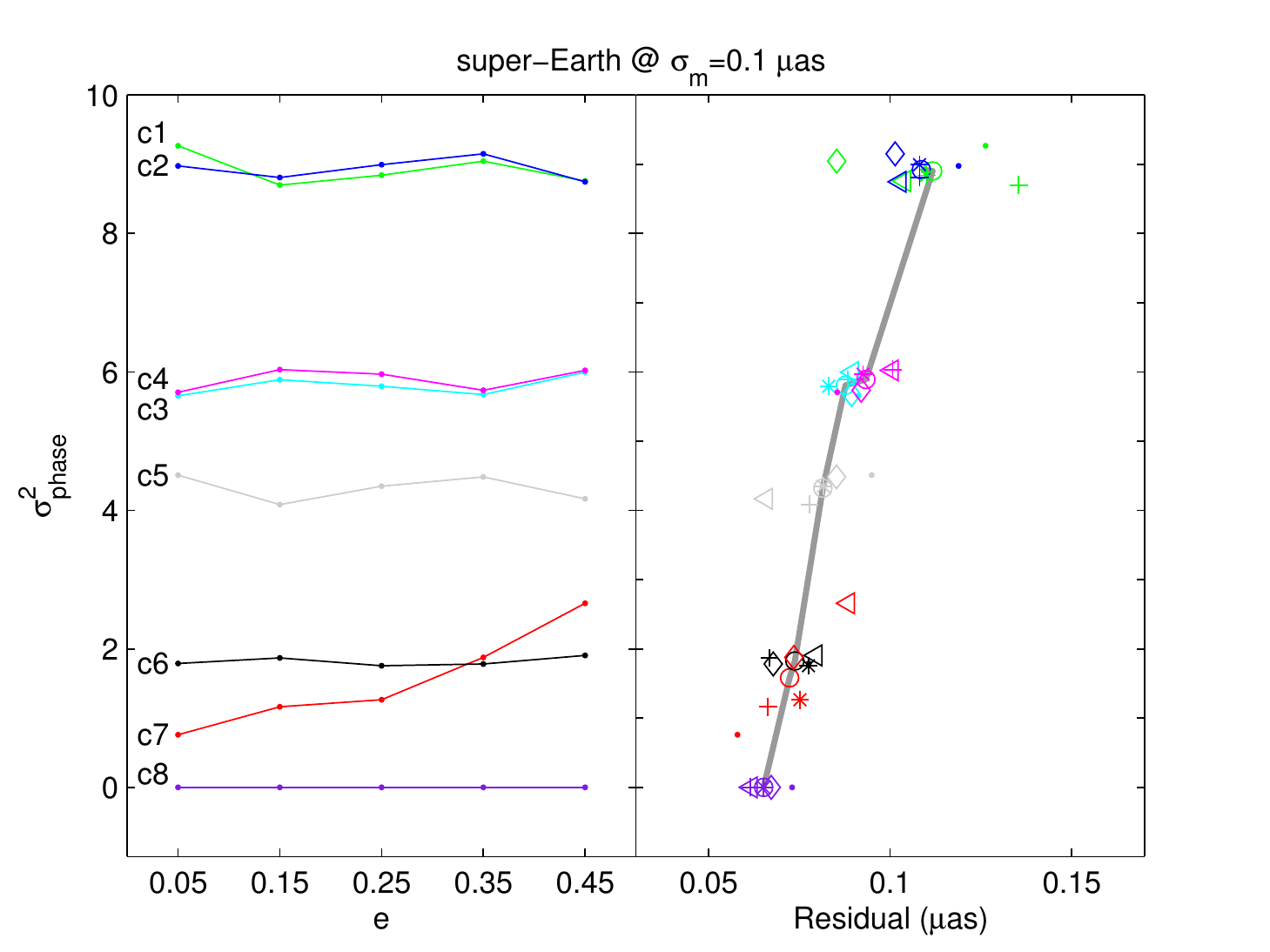}
\vspace{0cm}
\captionsetup{font={footnotesize}}
\caption{ Similar to Figure \ref{fig13} but for the super-Earth with observational error $\sigma_m=0.1$ $\mu$as. \label{fig14} }.
\end{figure}

To illustrate the non-uniformity of the data points, we divide the whole phase coverage of $f$ into 25 parts, each with a width of $14.4^\circ$. Then we count the number of data points in each parts and calculate the variance($\sigma^2_{\rm phase}$) of them. The variance represents the phase coverage of the observation, i.e., the smaller the variance is, the more complete the phase coverage is. For each kind of data cadence, the variance changes slowly with eccentricity, therefore, we calculate the average variance in each bin of eccentricities, the bin range is set to be 0.1. Set SNR $\sim$ 10, i.e, $\sigma_m=0.1$ $\mu$as for the super-Earth and $\sigma_m=3$ $\mu$as for the Jupiter, we fit the planet parameters with data sets c1-c8. The differences between the true and fitted astrometric signatures caused by the planets are shown in Figure \ref{fig13} and Figure \ref{fig14}. The residuals are expressed as $\sqrt{\sum_{i=1}^N((X_{\rm fit}(t_i)-X_{\rm ture}(t_i))^2+(Y_{\rm fit}(t_i)-Y_{\rm ture}(t_i))^2)/N}$. N=50 is the number of data points.

Similar characteristics for single Jupiter or super-Earth systems are obtained in our simulations, which is reasonable, because the simulations are done with similar SNR. The left panels of Figure \ref{fig13} and \ref{fig14} show the variance of each data cadence at different eccentricities of Jupiters and super-Earths, respectively. The right panels show the corresponding residuals at each variance. In left panels, $\sigma^2_{\rm phase}$ increases from c8 to c1. The cases of c8 have zero variances, while $\sigma^2_{\rm phase}$ for c6 and c7 are smaller than 3, which have much better phase coverage than c1-c5. $\sigma^2_{\rm phase}$ of c1 are similar to that of c2, because $f$ of data points near the perigee and apogee are uniformly distributed as shown in the top two panels in Figure \ref{fig12}. The same reason can also explain the similarity of $\sigma^2_{\rm phase}$ in c3 and c4. In the right panel of Figure \ref{fig13} and \ref{fig14}, with the increase of average variance, the residual also increases, indicating that more uniform and complete phase coverage will ensure a better orbit fitting of the planets.  With the similar $\sigma^2_{\rm phase}$ in c1 and c2, the residuals are nearly the same, i.e., the residuals are not sensitive to the samplings with more data near perigee or apogee. The similar residuals of c3 and c4 show there is no differences for data sampling near perigee/apogee or not. For even data cadence c7,  when eccentricities are larger, we'll have more data points near the apogee if we sample every 0.1yr, and the variance increases with the eccentricities. Accordingly, the increase of variance leads to the increase of residuals with the eccentricities, while there are no such obvious correlations for other cadences. Empirically, if $\sigma^2_{\rm phase}<3$, e.g. c6-c8, the residuals are smaller than observational errors $\sigma_m$ for single planet systems.

For single planet systems, even with extremely uneven data cadence, all the planets are detected with precise periods, although the residuals vary a lot. When it comes to two-planet systems, things are quite different. The large fitted residuals of the first planet may contaminate the signal of the secondary planet, thus the period of the secondary planet is hardly determined accurately. So we compare the differences between even and uneven data cadence for two-planet systems to see how large the influence is. As we adopt a keplerian orbit for each planet, the motion of the host star will be irregular rather than a keplerian orbit. So it's hard to clearly choose data points near the perigee or apogee for both planets. For two planet systems, the star moves around the common center of mass and locates in different quadrants at different times. Define $\alpha$ as the angle of data vectors [$x(t_{i})$, $y(t_{i})$](i=1,...N) with the x axis. We choose all the simulated astrometry data of super-Earth pairs in 2:3 MMR in section 4, and test 4 kinds of data samples as follows:
\begin{itemize}
\item[d1:] Sample in regions with $45^\circ<\alpha<90^\circ$ and $245^\circ<\alpha<270^\circ$.
\item[d2:] Sample in regions with $0^\circ<\alpha<90^\circ$ and $180^\circ<\alpha<270^\circ$.
\item[d3:] Sample in regions with $-45^\circ<\alpha<90^\circ$ and $135^\circ<\alpha<270^\circ$.
\item[d4:] Sample every 0.1 year, i. e., even data samples.
 \end{itemize}

 Table \ref{tab10} shows the results of the 4 kinds of sampling for super-Earth pairs in 2:3 MMRs with $\sigma_m=0.1$ $\mu$as. We choose $\sigma_m=0.1$ $\mu$as in order to ensure a large signal-to-noise ratio $\sim10$. Therefore, planets can be detected with large confidence and we can compare the influence of different sampling schemes on MMR-reconstruction in our simulations. Similar with single planet systems, we can calculate the variance $\sigma_{\rm phase,i}$ for each planet. From data cadence d1 to d4, the mean values of $\sigma^2_{\rm phase, i}$, denoted as $\overline{\sigma^2}_{\rm phase, i}$ (i=1,2), for both planets largely drop. The reason is obvious because the larger regions we sample in, the more uniform and complete phase coverage we'll have. When sampling only in a very small region, take d1 for example, only about $27\%$ of the results converge at $\chi_{\rm red}^2<1.5$, while for d3 and d4, all results can converge at small $\chi_{\rm red}^2<1.5$ as shown in the third row in Table \ref{tab10}. Among results with$\chi_{\rm red}^2<1.5$, the average MMR-reconstruction probabilities $\overline{\beta}$ also increase with the phase coverage in the fourth row. We investigate the fitting errors of eccentricities, which decrease from d1 to d4, and lead to the decrease of $\overline{\beta}$. $\overline{\beta}$ for d1 is smaller than others, because $\overline{\beta}$ becomes very small if the period of one planet is determined ambiguously. In our simulations, about $10\%$ of the fitted super-Earth pairs with small $\chi_{\rm red}^2$ have large fitting errors of semi-major axis for the secondary planet ($\delta_{a_1}>0.1$) while periods of both planets in d2-d4 are determined well with $\delta_{a_i}<0.05$(i=1,2) as shown in the fifth row in Table \ref{tab10}. For planet pairs with $\chi_{\rm red}^2>2$, which occurs only in d1 and d2, most of them are characterized with false periods of the secondary planets with $\delta_{a_1}>0.1$. Therefore, these planet pairs in 2:3 MMR can be hardly reconstructed. The mean values of $\beta$ are all $<0.08$ for planet pairs with $\chi_{\rm red}^2>2$. We define the average variance of the two planet $\overline{\sigma^2}_{\rm phase}=(\overline{\sigma^2}_{\rm phase, 1}+\overline{\sigma^2}_{\rm phase, 2})/2$. Consistent with single planet systems, if $\overline{\sigma^2}_{\rm phase}<3$, e.g. d2-d4, the MMR-reconstructed probabilities are much better than d1 with $\overline{\sigma^2}_{\rm phase}>4$.

\begin{table*}[htbp]
\begin{center}

\begin{tabular}{c|cccc}
\tableline\tableline
data cadence&d1&d2&d3&d4\\\hline
$\overline{\sigma^2}_{\rm phase, 1}$/$\overline{\sigma^2}_{\rm phase, 2}$&3.92/5.54&2.36/3.18&2.21/2.49&1.15/0.80\\\hline
fraction of $\chi^2<1.5$&$27.12\%$&$98.37\%$&$100\%$&$100\%$\\
$\overline{\beta}$ of $\chi^2<1.5$&0.63&0.72&0.75&0.77\\
fraction of $\delta_{a_1}>0.1$ $\&$ $\delta_{a_2}<0.05$ when $\chi^2<1.5$&$9.52\%$&0&0&0\\\hline
fraction of $\chi^2>2$&$60.26\%$&$1.40\%$&$0$&$0$\\
$\overline{\beta}$ of $\chi^2>2$&0.01&0.08&--&--\\
fraction of $\delta_{a_1}>0.1$ $\&$ $\delta_{a_2}<0.05$ when $\chi^2>2$ &$93.36\%$&$66.67\%$&--&--\\\hline

\end{tabular}
\footnotesize

\end{center}
\captionsetup{font={footnotesize}}
\caption{Results of uneven data cadence of the super-Earth pairs in 2:3 MMR \label{tab10}}

\begin{flushleft}
Note. d1-d4 represent the four kinds of data cadence for super-Earth pairs in 2:3 MMR in Section 7, $\overline{\sigma^2}_{\rm phase, i}$(i=1,2) is the mean value of $\sigma^2_{\rm phase}$ for each planet, $\overline{\beta}$ is the mean MMR-reconstruction probability for all super-Earth pairs in 2:3 MMR. Fraction of $\delta_{a_1}>0.1$ $\&$ $\delta_{a_2}<0.05$ when $\chi^2<1.5$(or $\chi^2>2$) represent the fraction of planet pairs with relative fitting errors of semi-major axis $\delta_{a_1}>0.1$ and $\delta_{a_2}<0.05$ among fitted planet pairs with $\chi^2<1.5$(or $\chi^2>2$).
\end{flushleft}
\end{table*}

The comparison between even and uneven data cadence indicates that it's important to have more uniformly distributed data points in astrometry measurements. Although even data cadence is hard to be carried out in real observations considering limitations of observational windows, we can obtain good MMR-reconstruction probabilities if the data sampling has a small variance $\overline{\sigma^2}_{\rm phase}<3$ according to our results. Choosing an even data cadence would be suitable for most cases except systems with very eccentric planet pairs in MMR, which are very rare.

\section{Conclusion and Discussion}

Astrometry is an ancient technique to discover asteroids and planets in solar system. With the improvements of technique, astrometry method can be extended to discover the exoplanets around nearby stars to obtain more information about the orbit of planets. Using these orbital elements and mass of star, we can reconstruct planet systems in mean motion resonances. In Section 2, we introduce the astrometry methods to detect planets and the fitting procedure of planetary parameters. Based on observations about planet pairs near MMRs(Figure \ref{fig1}), we consider planet pairs with equal masses, i.e., Jupiter pairs and super-Earth pairs. We also present how to simulate samples of planetary systems in 1:2, 2:3 and 3:4 MMRs via migration and random methods. Distribution of eccentricities and $\Delta$ for each MMR of the Jupiter pairs and super-Earth pairs are shown in Figure \ref{fig2}, Figure \ref{fig4} and Figure \ref{fig5}. In Section 3, planets with SNR$>3$ can be detected reliably (Figure \ref{fig3}) in our simulations. As we use the Keplerian orbit to model the true orbit, the difference may lead to false detection of third planets when there is no observational error, however, they can be ignored in our MMR-reconstruction because of their small masses and large separations with the detected planets.

In Section 4, we show the probabilities of reconstructing the Jupiter pairs and the super-Earth pairs in 1:2, 2:3 and 3:4 MMRs. The main conclusions are listed as follows:

1. The fitting errors of planet pairs are sensitive to observational errors according to Table \ref{tab11} and \ref{tab12}. The fitting errors lead to obvious decrease of the MMR-reconstruction probabilities $\beta$ with the decrease of SNR as shown in Table \ref{tab1} and \ref{tab2}.

2. With the increase of $\Delta$, there is a decrease in MMR-reconstruction probability $\beta$ for Jupiter pairs in 2:3 and 1:2 MMRs in Figure \ref{fig6}, which is not obvious for super-Earth pairs in Figure \ref{fig7}.

3. There is a positive correlation between MMR-reconstruction probability and the eccentricity of the planets for both Jupiter and super-Earth pairs in Figure \ref{fig8} and Figure \ref{fig9}. Planet pairs with $e_1>0.1$ and $e_2>0.1$ are better reconstructed than those with $e_1<0.1$ and $e_2<0.1$. Because large eccentricity can avoid the degeneracy between $\omega$ and $M$ and resonance width increase with eccentricity \citep{Deck2013}.

4. MMR-reconstruction probabilities are larger for planet pairs with strong resonance intensity with $A_{\phi_{i}}<30^\circ$(i=1,2) illustrated in Figure \ref{fig10} and Figure \ref{fig11}.

5. With similar SNR, the MMR-reconstruction probabilities of Jupiter pairs are larger than those of super-Earth pairs when considering stability, as shown in Table \ref{tab3}.

In Section 5, we calculate the FAPs when we reconstruct a planet system in or near MMRs. Our main conclusions are:

1. $P_{0-1}$ as the probability of mistaking a near MMR system for a resonant system has a positive correlation with observational error, meanwhile, it decreases with the increases of $\Delta_{\rm fit}$. The results are presented in Table \ref{tab4} and Table \ref{tab5}.

2. The FAPs for planets reconstructed to be in MMR $\rm \mathscr{F}_{0-1}$ are largest for planet pairs in 1:2 MMR. It's difficult to produce a stable Jupiter pair near 3:4 MMR, thus $\rm \mathscr{F}_{1-0}\sim0$. Both $\rm \mathscr{F}_{0-1}$ and $\rm \mathscr{F}_{1-0}$ are sensitive to observational errors. As shown in Table \ref{tab7} and Table \ref{tab8}, when SNR $\sim$ 3, both $\rm \mathscr{F}_{0-1}$ and $\rm \mathscr{F}_{1-0}$ are larger than $30\%$, so planets with small SNR detected to be in MMRs should be checked carefully.

 In Section 6, we estimate the number of discovering planet systems in MMRs via astrometry, as shown in Table \ref{tab9}. There are about $3\times10^4$ stars with V$<$10 within 30 pc from the Sun, after assuming the occurrence of planet pairs in MMRs, we estimate that with SNR$=3$, tens of planet pairs with Jupiter masses in 2:3, 1:2 and 3:4 MMRs can be potentially reconstructed, and hundreds of super-Earth pairs in 2:3 and 1:2 MMRs can be detected, planet pairs in 3:4 MMRs are very few because of their rareness based on observation.

In Section 7, we compare the difference between even and uneven data cadences. Extremely uneven data cadence with $\sigma^2_{\rm phase}>4$ leads to large fitting errors in single planet systems, while data cadence with good phase coverage with $\sigma^2_{\rm phase}<3$ have good fitting results(see Figure \ref{fig13} and \ref{fig14}). Although it's hard to have even data cadence in real observations, it's important to have enough data points to guarantee a good phase coverage. Using a defined parameter $\overline{\sigma^2}_{\rm phase}$ in two planet systems, the MMR-reconstruction probabilities with $\overline{\sigma^2}_{\rm phase}<3$ are similiar with even data cadence (see Table \ref{tab10}).

Nowadays, the precision of the GAIA program is about a few tens of $\mu$as, which can help us find planets of Jupiter mass. If it can reach a precision of about 10 $\mu$as, the probabilities to reconstruct a Jupiter pair in 2:3 and 1:2 MMRs $>50\%$ at least (see Table \ref{tab1}). If a Jupiter pair with such an SNR is reconstructed in MMR, it should be checked very carefully because of the large FAP $\sim40\%$. The target precision of the STEP program for bright stars is about 1 $\mu$as, for a super-Earth 1 AU from the host star and 30 pc from us, the SNR $\sim1$, which is very hard to identify the super-Earth. However, if the planets is 10 pc from us, the SNR $\sim3$, which will ensure a probability of $40\%$ with FAP $\sim40\%$ for 2:3 and 1:2 MMR. We expect higher precision of astrometry($\sim0.1$ $\mu$as) in the future, thus we will have chances to detect planets with masses even smaller than Earth, and the probability to reconstruct super-Earth pairs in MMRs will be improved to as large as $75\%$(Table \ref{tab2}). All planet systems in our simulations are at 30 pc, with similar observational errors, we can reconstruct planet pairs in MMRs with larger probability and smaller FAPs if they are closer to us.

In this paper, we adopt a mission lifetime of 5 years comparable with GAIA and STEP. Thus it's appropriate to detect planet systems around 1 AU from the host star. Data cadence and the time allocated for observations will influence our planet detecting via astrometry. Shorter data cadence helps us to detect planets closer to the host star, longer mission lifetime enables us to detect planets with longer period. Only 50 data are used in this paper, more data can improve the fitting precision thus may leads to larger MMR-reconstruction probability. Besides, recent work \citep{Giuppone2009,Giuppone2012} have shown potential of detecting and characterizing planet pairs in MMRs using radial velocity data, together with high precision radial velocity data, we can improve the precision of eccentricities which can help us determine $\omega+\Omega$ accurately. Consequently, planet pairs in MMR can be reconstructed with large probabilities and small FAPs. Additionally, although fitting errors for planets in 1:2 MMRs are larger than those in 2:3 and 3:4 MMRs, it's hard to conclude whether planet pairs in 1:2 MMR are harder to be reconstructed than the other two MMRs or not according to results of our simulations, because many factors influence the MMR-reconstruction probabilities. E.g., different MMRs have different resonance width and resonance structures in $e_1-e_2$ phase diagram, it's hard to have a large number of samples with exactly same distribution of $\Delta$, eccentricities and amplitudes of resonance angles for different MMRs. The same reason fits to comparison between super-Earth pairs and Jupiter pairs, which are also difficult to figure out the significant differences between them.

We only simulate planet pairs with equal masses in the first order MMRs in this paper, other planet systems in MMRs with different masses such as a Jupiter and a super-Earth can also be reconstructed with proper observational precision. Planet pairs in high order MMRs such as 1:3 and 3:5 are not considered here, as these MMRs are much weaker and have a narrower resonance width than the MMRs discussed in this paper, they need higher precision to be reconstructed.

This research is supported by the Key Development Program of Basic Research of China (No. 2013CB834900), the National Natural Science Foundations of China (Nos. 11503009, 11003010 and 11333002), Strategic Priority Research Program The Emergence of Cosmological Structures of the Chinese Academy of Sciences (Grant No. XDB09000000), the Natural Science Foundation for the Youth of Jiangsu Province (No. BK20130547), 985 Project of Ministration of Education and Superiority Discipline Construction Project of Jiangsu Province, "Search for Terrestrial Exo-Planets", the Strategic Priority Research Program on Space Science Chinese Academy of Sciences( Grant No. XDA04060900).

\addtolength{\bibsep}{10pt}

\clearpage

\clearpage

\appendix

\section{Appendix information about the MCMC procedure}

\begin{figure}
\vspace{0cm}\hspace{0cm}
\centering
\includegraphics[scale=0.9]{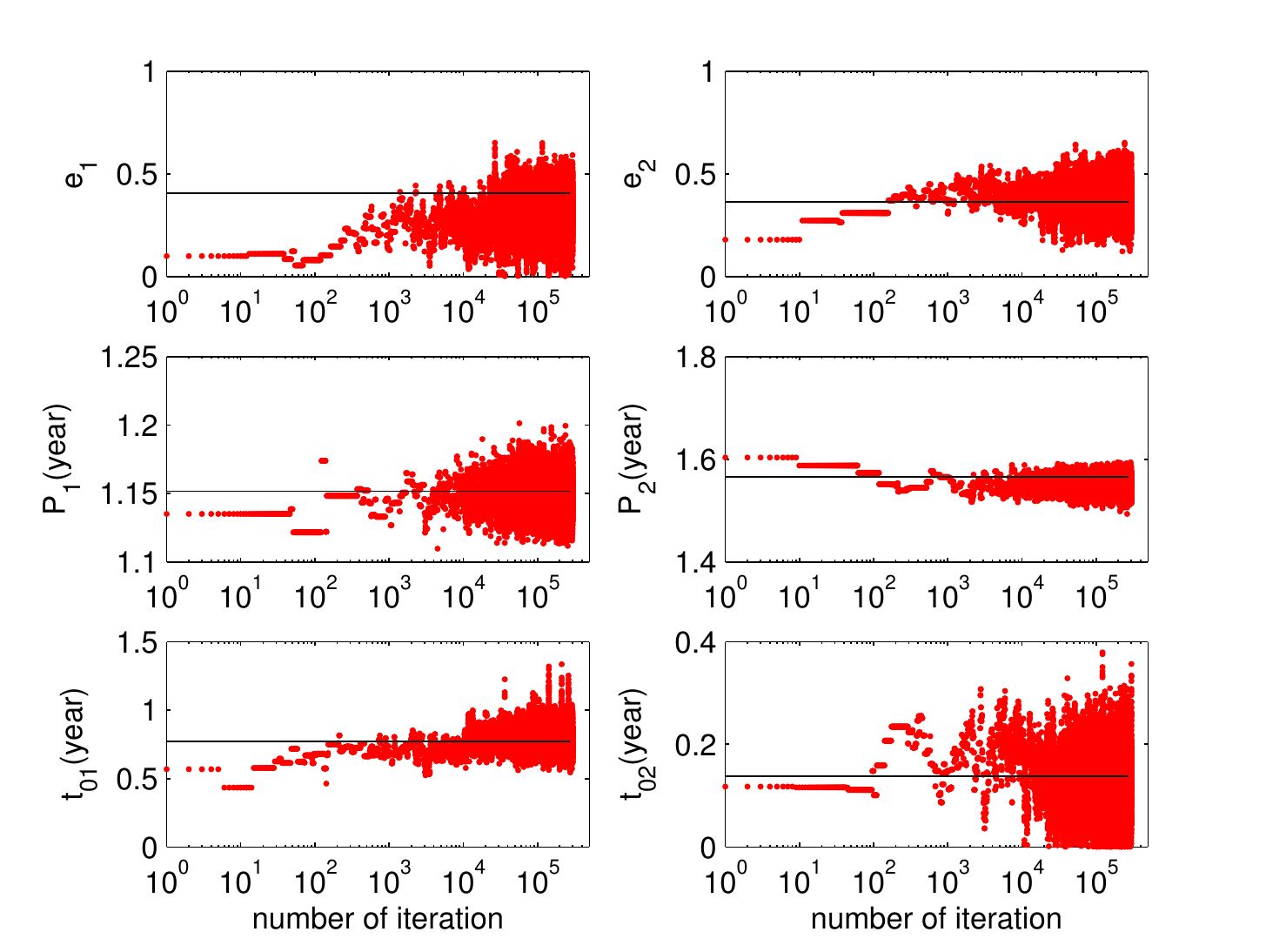}
\vspace{0cm}
\caption{The trace plot of the iteration number against the value of the parameters at each iteration for a Jupiter pair in 3:4 MMR with $\sigma_m=10$ $\mu$as. The dark lines are the true values of each parameter of the Jupiter pair.\label{figa1} }.
\end{figure}

\begin{figure}
\vspace{0cm}\hspace{0cm}
\centering
\includegraphics[scale=0.9]{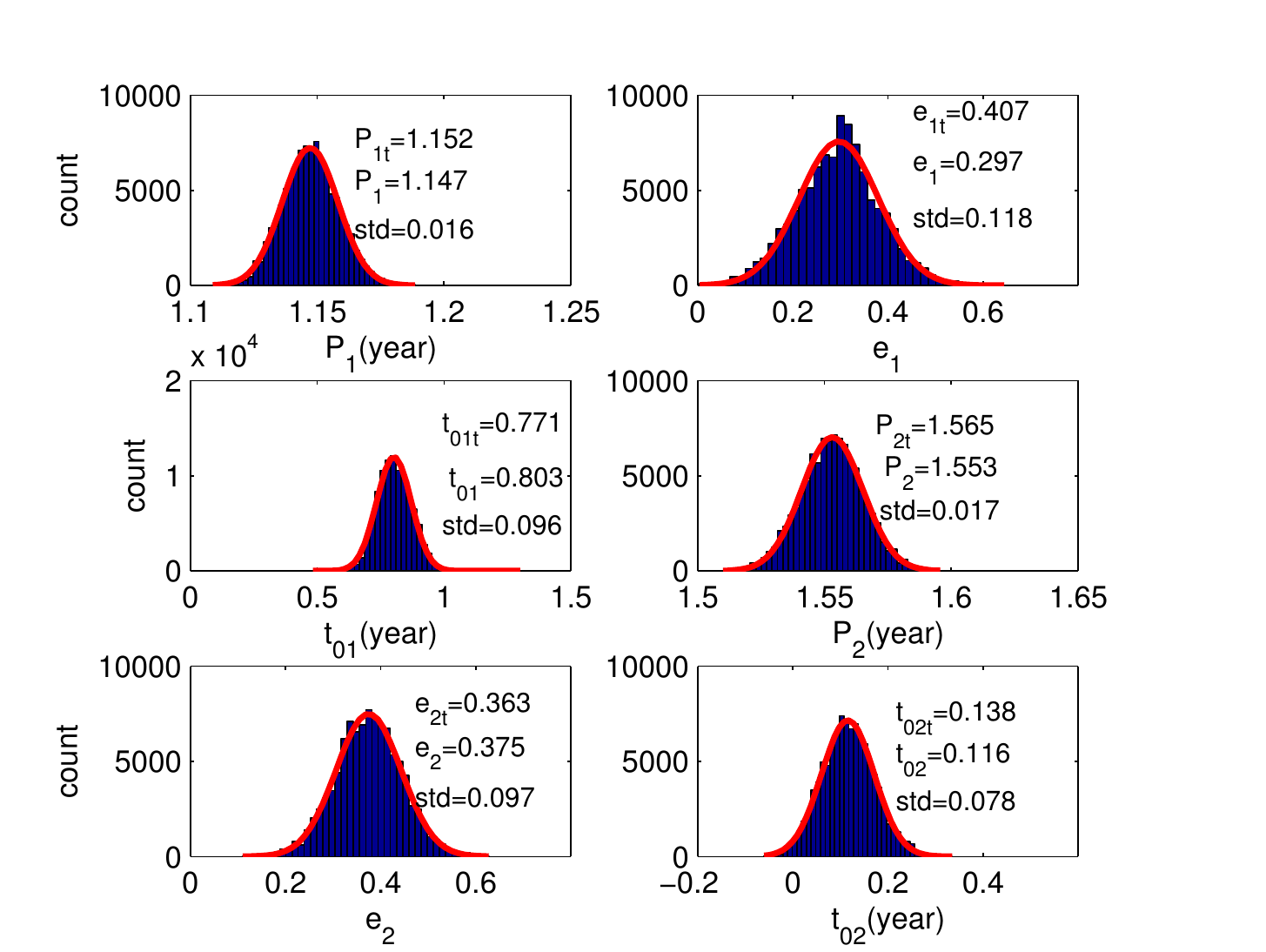}
\vspace{0cm}
\caption{ The distribution of the last $1\times10^5$ fitted parameters of a Jupiter pair in 3:4 MMR with observational errors(blue) $\sigma_m=10$ $\mu$as. The red lines are the Gaussian fit to the distribution of the parameters. $P_{1t}$, $e_{1t}$, $t_{01t}$, $P_{2t}$, $e_{2t}$ and $t_{02t}$ are the true values while $P_1$, $e_1$, $t_{01}$, $P_2$, $e_2$ and $t_{02}$ are the median value of the Gaussian fit, $std$ represent the corresponding standard deviations of the fitted parameters. \label{figa2} }.
\end{figure}

\begin{figure}
\vspace{0cm}\hspace{0cm}
\centering
\includegraphics[scale=0.9]{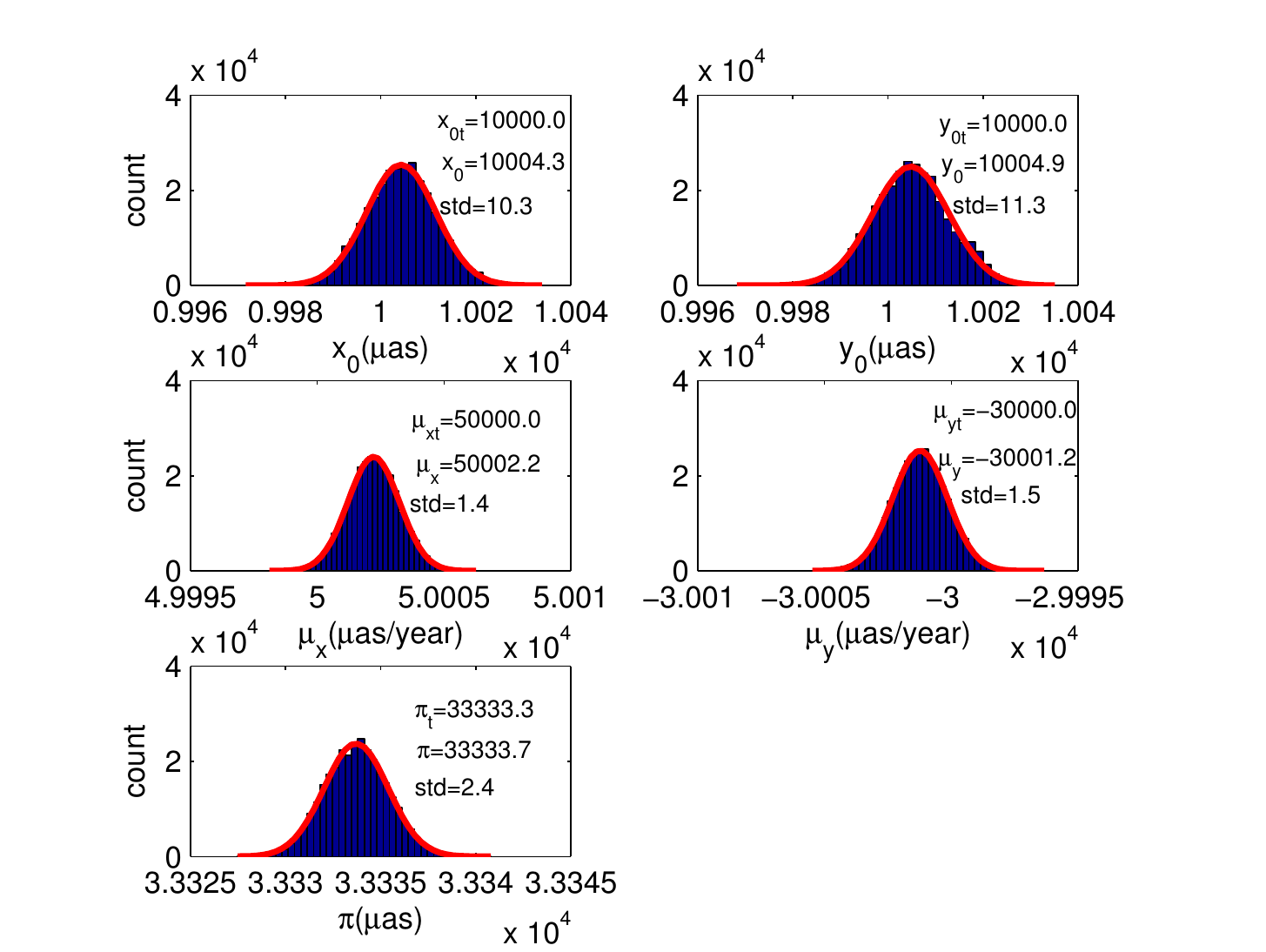}
\vspace{0cm}
\caption{ The distribution of the last $1\times10^5$ fitted stellar parameters of a Jupiter pair in 3:4 MMR with observational errors $\sigma_m=10$ $\mu$as(blue). The red lines are the Gaussian fit to the distribution of the parameters. $x_{0t}$, $y_{0t}$, $\mu_{xt}$, $\mu_{yt}$ and $\pi_t$ are the true values while $x_0$, $y_0$, $\mu_x$, $\mu_y$ and $\pi$  are the median value of the Gaussian fit, $std$ represent the corresponding standard deviations of the fitted parameters. \label{figa3} }.
\end{figure}

The MCMC code is written based on theories discussed in \citet{Ford2005,Ford2006}. We run each MCMC with $3\times10^{5}$ iterations. First of all, we check the values of the parameters at each iteration in the MCMC procedure and find that each of them converges to a small range near the true values after $10^{4}$ iterations (Figure \ref{figa1}). Secondly, the acceptance of each parameter is around 0.2-0.5 and $\chi_{\rm red}^2<1.7$, we then conclude that the Markov chains are convergent. Statistics are derived on the last $1\times10^5$ elements, which can reduce the dependence on the initial parameter values. We choose the best-fit parameters as the median of posterior distribution. Figure \ref{figa2} and Figure \ref{figa3} show the posterior distributions of planetary orbital parameters and the stellar parameters of one case in our simulation. We can see that distributions fit well to Gaussian distribution and the true values of the parameters locate within one sigma range of the median values.

\begin{figure}
\vspace{0cm}\hspace{0cm}
\centering
\includegraphics[scale=0.9]{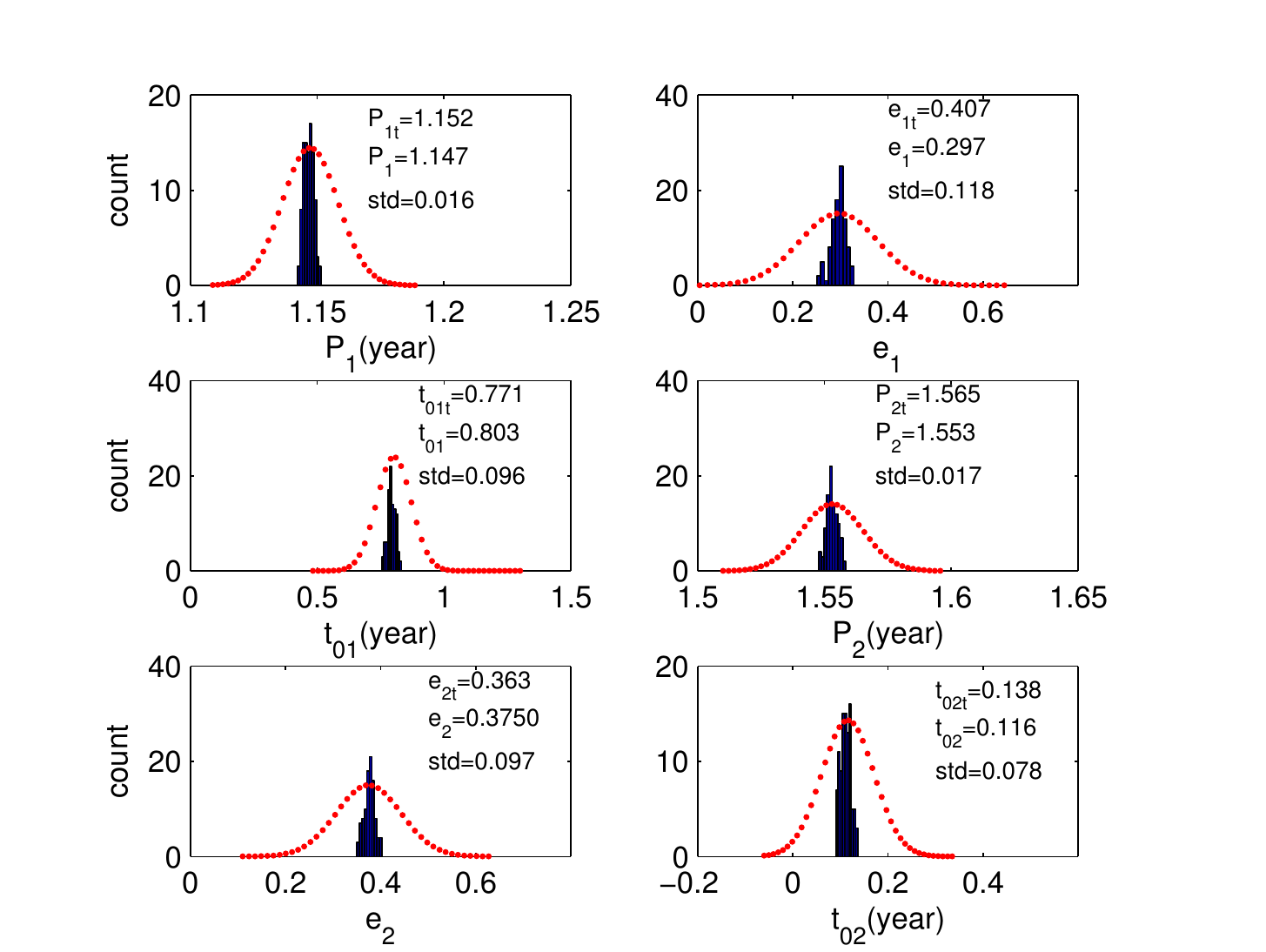}
\vspace{0cm}
\caption{ The distribution of the values for the best fit parameters of a Jupiter pair in 3:4 MMR with observational errors $\sigma_m=10$ $\mu$as(blue)(the same with that in Figure \ref{figa1}) with 100 different initial values of $e_1$, $t_{01}$, $e_{2}$ and $t_{02}$. The red lines are the Gaussian fit lines in Figure \ref{figa2}. In order to have a clear comparison, the Gaussian fit values are 500 times smaller than those in Figure \ref{figa2}. $P_{1t}$, $e_{1t}$, $t_{01t}$, $P_{2t}$, $e_{2t}$ and $t_{02t}$ are the true values while $P_1$, $e_1$, $t_{01}$, $P_2$, $e_2$ and $t_{02}$ are the median value of the Gaussian fit in Figure \ref{figa2}. \label{figa4} }.
\end{figure}

Generally, when we need to fit a lot of parameters using the MCMC method, we should have as many as iterations we can or have enough chains to approach a global best solution. However, among the 11 parameters we need to fit, the initial values of the 5 stellar parameters in our procedure are derived by the linear least squares, and the initial periods of the planets are derived by the periodogram. Stellar parameters and period of planets derived here are very close to the true values. Therefore, we only need to set initial values for 4 parameters $e_1$, $t_{01}$, $e_2$ and $t_{02}$ randomly. During the fitting procedure, $t_{0i}$ is set to change between $0$ and $P_{i}$(i=1,2), while $e_{i}$ is set to change between $0$ and $1$. With a limited parameter space, the MCMC procedure is more efficient \citep{Ford2007}. $3\times10^5$ iterations are enough to lead a high confidence that we have reached a global solution. To confirm that, we simulate the same case in Figure \ref{figa1} with 100 different initial values of $e_1$, $t_{01}$, $e_2$ and $t_{02}$. The iteration number is $3\times10^5$. The median values of the posterior distribution are shown in Figure \ref{figa4}. We can see that the median values of the 100 chains locates within one sigma range of the median values shown in Figure \ref{figa2}. Therefore, we conclude that the chains are convergent with iterations of $3\times10^5$ and different initial values lead to similar results.

\clearpage

\end{document}